\documentclass{aa}  

\usepackage{graphicx}
\usepackage{txfonts}
\usepackage{xcolor}
\usepackage{ulem}
\usepackage{tikz}
\newcommand{\beqa}{\begin{eqnarray}}
\newcommand{\eeqa}{\end{eqnarray}}
\newcommand{\bea}{\begin{eqnarray}}
\newcommand{\eea}{\end{eqnarray}}
\usepackage{amstext}

\begin{document}



\title{Cosmic voids in modified gravity scenarios}

\author{
Eder L. D. Perico\inst{1,2}
\and Rodrigo Voivodic\inst{2}
\and Marcos Lima\inst{2}
\and David F. Mota\inst{1}}
\institute{
Institute of Theoretical Astrophysics, University of Oslo, PO Box 1029 Blindern, N-0315 Oslo, Norway
\and
Departamento de F\'isica Matem\'atica, Instituto de F\'isica, Universidade de S\~ao Paulo, Rua do Mat\~ao 1371, S\~ao Paulo, SP, 05508-090, Brazil
}

\offprints{E. L. D. Perico, \email{elduartep@usp.br}}

\date{Received d M y / Accepted d M y}

\abstract{
Modified gravity (MG) theories aim to reproduce the observed acceleration of the Universe by reducing the dark sector while simultaneously recovering General Relativity (GR) within dense environments.
Void studies appear to be a suitable scenario to search for imprints of alternative gravity models on cosmological scales.
Voids cover an interesting range of density scales where screening mechanisms fade out, which reaches from a density contrast $\delta \approx -1$ close to their centers to $\delta \approx 0$ close to their boundaries.
We present an analysis of the level of distinction between GR and two modified gravity theories, the Hu-Sawicki $f(R)$ and the symmetron theory.
This study relies on the abundance, linear bias, and density profile of voids detected in n-body cosmological simulations.
We define voids as connected regions made up of the union of spheres with a {\it \textup{mean}} density given by $\overline\rho_v=0.2\,\overline\rho_m$, but disconnected from any other voids.
We find that the height of void walls is considerably affected by the gravitational theory, such that it increases for stronger gravity modifications.
Finally, we show that at the level of dark matter n-body simulations, our constraints allow us to distinguish between GR and MG models with $|f_{R0}| > 10^{-6}$ and $z_{SSB} > 1$.
Differences of best-fit values for MG parameters that are derived independently from multiple void probes may indicate an incorrect MG model. This serves as an important consistency check.
}


\maketitle 



\section{Introduction}
\label{sec:intro}

Based on Einstein's General Relativity (GR) theory, most of the observational features of the Universe on cosmological scales are nicely reproduced by the so-called cosmic concordance model or $\Lambda$CDM \citep{Perlmuter03, BAO, Kowalski_2008,x-ray08, wmap_2013, planck_2018}.
Nevertheless, one of the main problems in modern cosmology is understanding the nature of dark energy. This is the exotic energy component with negative pressure that is required by the standard $\Lambda$CDM model to reproduce the accelerated expansion of the Universe, which is observed at low redshift \citep{RiessETAJ98, PerlmuterETNature98}.
The simplest candidate for dark energy is the cosmological constant, denoted by $\Lambda$, which is a free geometrical parameter of Einstein's theory.

Some alternatives to the cosmological constant are the quintessence or dark sector interaction models, for instance, and running the vacuum expectation value.
For a review, see \cite{Yoo:2012ug}, \cite{Joyce2015}, and references therein.
We are here interested in another set of theories, that is, theories of modified gravity (MG). These aim to model the accelerated expansion of the late Universe by going beyond GR.
Viable MG theories must display a screening mechanism, which consists of the weakening of the gravity modifications within dense environments, such as our Solar System, where GR has been exhaustively tested.
Different families of MG theories are characterized by different ways to accomplish the screening effect \citep[see, e.g.,][]{PhysRevD.86.044015,Koyama16}.

When efficient screening mechanisms are available, MG theories are virtually indistinguishable from GR inside massive halos.
On the other hand, long-range forces as well as cumulative effects due to different time-evolution paths can modify the large-scale spatial distribution and the abundance of halos.
These observables have recently been studied in different contexts \citep[e.g.,][]{SchmidtETPRD09,n-body-comparisonMNRAS15, Koyama16}.
In contrast, we here focus on a complementary scenario where screening mechanisms could be weak, leading to a detectable signature of modified gravity.
Our goal is to test MG through the analysis of cosmic voids, which are the intermediate-sized and underdense regions that are left behind by the hierarchical structure formation of dark matter halos.

By definition, voids are underdense regions naturally bounded by an overdense or a background-dense wall.
 Methods to detect them include free-parameter geometrical algorithms intended to find non-spherical voids \citep{Lavaux_2012,10.1111/j.1365-2966.2008.13180.x,watershedMNRAS07}.  Spherical void finders are based on theoretical density thresholds \citep{10.1111/j.1365-2966.2008.13307.x}.

Many cosmological probes based on void analysis have been proposed in the past decade in the context of GR, including a number of observables such as ellipticity \citep{voids-elipticityPRL07,elipticity-voidsArxiv12}, density profile \citep{SDSS-universal-void-profile14,DESvoids17,Nadathur:2019mct}, and gravitational lensing \citep{10.1093/mnras/stw2745,Krause_2012}.
Moreover, void properties depend strongly on the void-finding algorithm \citep{watershedMNRAS07,zobov-voidsRD14,10.1093/mnras/stv043,stacking-voidsMNRAS16,magneticum-simulation-voidsMNRAS17}.
Despite this dependence on the void-finding algorithm, the void analysis can be performed in a self-consistent way by calibrating the methods on mock catalogs, which plays the role of the theory, and 
this has been successfully applied to photometric samples \citep{Nadathur:2019mct} that improved the Baryonic Acoustic Oscillations (BAO) constraints.
This type of analysis shows great promise if it were applied to future experiments, including DESI \citep{DESI} and Euclid \citep{EUCLID}.

The study of voids in the context of modified gravity has recently gained strength.
This includes mainly the effect of MG on the void abundance \citep{voids-halosMNRAS12,JenningsLiHuMNRAS13,MGvoidsMNRAS15,halo-voidsMNRAS15,RodrigoEtPRD17} and in the lensing signature around voids \citep{Barreira_2015,PhysRevD.92.064005,PhysRevD.98.023511}.
In these scenarios, the galaxy distribution is modified by the fifth force associated with MG, and in turn, this also affects the shape and abundance of voids \citep{galaxy-voidsMNRAS15}.
MG theories must reproduce the Newtonian gravitational force within dense regions while displaying the characteristic repulsive effect of the cosmological constant in background density environments.
The transition between these two asymptotic behaviors is a potential scenario to probe MG models.
We therefore focus here on void-related observables, and compare the effects of different screening mechanism effects on the $\Lambda$CDM outcome.

Specifically, we aim to distinguish between the $\Lambda$CDM model and two MG theories.
The first theory is the Hu-Sawicki $f(R)$ theory \citep{Hu-SawickiPRD07}, which implements the chameleon screening mechanism.
The second theory is the symmetron model \citep{symmetronPRD08,symmetronPRL10,symmetronPRD11}, which implements the screening mechanism that bears its name.

This study was carried out in the context of an idealized scenario, at $z=0$ and with a fixed set of cosmological parameters whose variation would lead to already observed degeneracies.
Generalizations, like those cited above, will be part of our future efforts.

The paper is organized as follows.
In section \ref{sec:models} we give an overview of the MG models we are interested in.
In section \ref{sec:meth} we describe the n-body simulations we analyze in this work, as well as the void-finder algorithm we employ.
In section \ref{sec:observables} we describe the three void-related observables we choose to analyze, how we estimate them from the n-body simulations, and the phenomenological fitting expressions we use to describe them.
In sections \ref{sec:right_analysis} and \ref{sec:using_wrong_model} we show how well we can distinguish among GR and the two MG theories based on the abundance, linear bias, and density profile of voids.
Finally, in section \ref{sec:conclusions} we present our conclusions.

\section{Gravity models}
\label{sec:models}

\subsection{Symmetron}
\label{subsec:symmetron}

The symmetron cosmological model \citep{symmetronPRL10,symmetronPRD11,symmetronPRD08} is a scalar-tensor theory for a single scalar field $\phi$.
The action for a scalar-tensor theory can be written as
\begin{equation}\label{eq:symmetron_action}
S=\int d^4x\sqrt{-g}\left[\frac{R}{2}M_{pl}^2-\frac{\nabla_\mu\phi\nabla^\mu\phi}{2}-V(\phi) \right]+S_m(\tilde g_{\mu\nu},\psi_i)\,,
\end{equation}
where $S_m$ is the action associated with the standard matter fields $\psi_i$.
The fields $\psi_i$ are coupled to the scalar field $\phi$ through the Jordan frame metric $\tilde g_{\mu\nu}$, which is related to the Einstein frame metric $g_{\mu\nu}$ by $\tilde g_{\mu\nu}=A^2(\phi)g_{\mu\nu}$.
For the particular case of the symmetron theory, we have
\begin{equation}\label{eq:Jordan-Einstein_factor}
A(\phi)=1+\frac{1}{2}\left(\frac{\phi}{M} \right)^2\,,
\end{equation}
and
\begin{equation}\label{eq:symmetron_potential}
V(\phi)=V_0-\frac{\mu^2\phi^2}{2}+\frac{\lambda\phi^4}{4}\,.
\end{equation}

In this case, both $A(\phi)$ and $V(\phi)$ are symmetric upon the transformation $\phi\rightarrow-\phi$.
Here, $\mu$ and $M$ are mass scales, $\lambda$ is a dimensionless coupling constant, $M_{pl}^2=(8\pi G)^{-1}$ is the Planck mass scale, and $g$ is the determinant of $g_{\mu\nu}$.

The variation of the action in Eq.~\eqref{eq:symmetron_action}, with respect to the Jordan frame metric provides the dynamical equation for the scalar field $\phi$:
\begin{equation}\label{eq:symmetron_effective_potential}
\Box\phi=V_{\text{eff}}=\frac{1}{2}\left(\frac{\rho_m}{\mu^2} -M^2\right)\phi^2+\frac{\lambda\phi^4}{4}\,.
\end{equation}

The effective potential $V_\text{eff}$, Eq. \eqref{eq:symmetron_effective_potential}, has a minimum at $\phi=0$ in high-density environments, that is, where the local matter density $\rho_m\gg\mu^2 M^2\equiv\rho_{SSB}$.
In this case, $V_\text{eff}$ holds the $\phi\rightarrow-\phi$ symmetry.
On the other hand, in low-density environments, $\rho_m\ll\rho_{SSB}$, the effective potential displays two minima at $\phi=\pm\phi_0\sqrt{1-\rho_m/\rho_{SSB}}$, breaking the symmetry of the model.
Here $\phi_0=\mu/\sqrt{\lambda}$ is the expected value of $\phi$ for $\rho_m=0$.

The free parameters of the symmetron model $\{\mu,\,M,\,\lambda\}$ are usually exchanged by the physical parameters associated with the scalar field for $\rho_m=0$ \citep{WintherMotaLiApJ12}:
\begin{equation}
\lambda_0=\frac{1}{\sqrt{2}\mu}\,,\qquad
\beta=\frac{\phi_0 M_{pl}}{M^2}\,,\qquad
(1+z_{SSB})^3=\frac{\mu^2M^2}{\overline\rho_{m0}}\,,
\end{equation}
which correspond to the range of the scalar field in Mpc$/h$ units ($\lambda_0$), the dimensionless coupling strength to matter ($\beta$), and the redshift for which the symmetry breaking occurs at the background level ($z_{SSB}), $ respectively.
$z_{SSB}$ is related to the symmetron critical density $\rho_{SSB}$, for which the symmetry breaking takes place through the expression $\rho_{SSB}=\overline\rho_{m0}(1+z_{SSB})^3$.
Here $\overline\rho_{m0}$ is the matter background density at redshift zero.

\subsection{f(R) Gravity}
\label{subsec:fr}

The Hu-Sawicki $f(R)$ model \citep{Hu-SawickiPRD07} was first formulated in the Jordan frame in terms of the action
\begin{equation}
S=\int d^4x \sqrt{-\tilde g} \left[\frac{\tilde R+f(\tilde R)}{16 \pi G} +\mathcal{L}_m\right]\,,
\end{equation}
where $\mathcal{L}_m$ is the Lagrangian describing the matter fields, and
\begin{equation}
f(\tilde R)\equiv -\frac{m^2c_1(\tilde R/m^2)^n}{c_2(\tilde R/m^2)^n+1}\,.
\end{equation}

In this case, $m^2=H_0^2\Omega_{m0}$, where $H_0$ and $\Omega_{m0}$ are the Hubble constant and matter density relative to the critical value today.
In order to recover the effective cosmological constant in the large curvature regime, it must be set $c_1/c_2=6\Omega_\Lambda/\Omega_m$.

This $f(R)$ theory can be transformed into a scalar-tensor theory upon both the identification
\begin{equation}
f_R\equiv\frac{df(\tilde R)}{d\tilde R}=e^{-2\beta\phi/M_{pl}}-1\approx-\frac{2\beta\phi}{M_{pl}}\,,
\end{equation}
and the frame transformation
\begin{equation}
\tilde g_{\mu\nu}=e^{2\beta\phi/M_{pl}}g_{\mu\nu}\,,\qquad\text{with}\qquad\beta=\frac{1}{\sqrt{6}}\,.
\end{equation}
In this case, the Compton wavelength or range of propagation of the scalar field at redshift zero is given by
\begin{equation}\label{eq:lambda0_fR}
\lambda_0=3\sqrt{\frac{n+1}{\Omega_m+4\Omega_\Lambda}}\sqrt{\frac{|f_{R0}|}{10^{-6}}}\frac{\text{Mpc}}{h}\,,
\end{equation}
where $f_{R0}$ can be expressed as a function of $\{c_2,\,n \}$ as
\begin{equation}
f_{R0}\equiv \left.f_R\right|_{z=0}=-\frac{6n\Omega_\Lambda}{c_2\Omega_m}\left(\frac{\Omega_m/3}{\Omega_m+4\Omega_\Lambda} \right)^{n+1}\,.
\end{equation}

\section{Methods}
\label{sec:meth}

\subsection{N-body simulations}
\label{subsec:simulations}

The cosmological n-body simulations analyzed in this work were run with the ISIS code \citep{isisAA14,llinares2}, which is a modification of the GR RAMSES code \citep{ramsesAA02} to include MG models, with $512^3$ dark matter particles in a $(256\,\text{Mpc}/h)^3$ cubic box.
The initial conditions correspond to a flat $\Lambda$CDM cosmology with parameters $\{\,\Omega_c\,,\,\Omega_b\,,\,h\,,\,\sigma_8\,,\,n_s\,\} = \{0.222\,,\,0.045\,,\,0.719\,,\,0.8\,,\,1\,\}$ and without neutrinos.
Massive neutrinos and modified gravity degeneracy has been pointed out and analyzed before \citep[see][]{Hagstotz:2019gsv,Schuster:2019hyl,10.1093/mnras/stz1944}, but we here prefer to exclude massive neutrinos to single out the modified gravity effects.

We intend to analyze void properties within the framework of the symmetron and f(R) gravity. N-body simulations are highly computationally expensive, therefore it is not feasible to cover the full parameter space of these theories.
On the other hand, we aim to show the MG effects on a recently explored void probe as clearly as possible, therefore we do not restrict our analysis to tighter constraints \citep{PhysRevD.99.043539,PhysRevD.77.023507,PhysRevD.81.049901}.
The MG cases we cover here are described by the parameters  $n=1$ and $|f_{R0}|=\{10^{-4},\,10^{-5},\,10^{-6}\}$ for the $f(R)$ theory, and by $\beta=1$, $\lambda_0=1~\text{Mpc}/h$ and $z_{SSB}=\{1,\,2\,,3 \}$ for the symmetron theory (see Tables~\ref{tab:symmetron_table} and \ref{tab:f(R)_table}).
The set of simulations is complemented by the $\Lambda$CDM or GR case, where no gravity modifications are included.
We assume here that the $\Lambda$CDM case corresponds to both the $|f_{R0}|=0$ and the $z_{SSB}=0$ scenarios.

Even though the effect of $f(R)$ and symmetron gravities may change over redshift, the effect is expected to be stronger at $z=0$.
In particular, symmetron and GR must be indistinguishable for $z>z_{SSB}$, while in linear theory the enhancement of gravity due to $f(R)$ weakens for higher redshift \citep{SchmidtETPRD09}.
We here only analyze the $z=0$ outputs and leave the redshift analysis for future works.

\begin{table}
        \centering
        \caption{Symmetron simulations analyzed in this work. All cases have $\lambda_0=1$ Mpc/$h$ and $\beta=1$.}
        \label{tab:symmetron_table}
        \begin{tabular}{ccccc}
                \hline\hline
                Symmetron case & $\Lambda$CDM & A & B & D\\
                \hline
                $z_{SSB}$ & 0 & 1 & 2 & 3\\
                \hline
        \end{tabular}
\end{table}

\begin{table}
        \centering
        \caption{$f(R)$ simulations analyzed in this work. All cases have $n=1$ and $\beta=1/\sqrt{6}$ , and $\lambda_0=\lambda_0(|f_{R0}|)$ is given by Eq. \eqref{eq:lambda0_fR} in Mpc/$h$ units.}
        \label{tab:f(R)_table}
        \begin{tabular}{ccccc}
                \hline\hline
                $f(R)$ case & $\Lambda$CDM & f6 & f5 & f4\\
                \hline
                $|f_{R0}|$ & 0 & $10^{-6}$ & $10^{-5}$ & $10^{-4}$\\
                $\lambda_0(|f_{R0}|)$ & 0 & 2.4 & 7.5  & 23.7\\
                \hline
        \end{tabular}
\end{table}

\subsection{Void-finding algorithm}
\label{subsec:void_finder}

Non-spherical voids, such as those found by methods based on Voronoi tessellation, display a dependency of the inner density on void size.
In these cases, the smaller the void, the emptier \citep{profilesPRL14}, in contrast to the spherical model prediction, where each void has the same mean density of about $0.2$ times the background density.
Because denser voids can naturally be larger, they leave less room for small and emptier voids. The abundance functions of voids that are detected through Voronoi tessellation and fixed density are therefore considerably different.

On the other hand, the ellipticity distribution of non-spherical voids also depends on the void size  \citep{voids-elipticityPRL07,elipticity-voidsArxiv12}.
Using the non-spherical version of our void-finder algorithm, which we describe in the next paragraph, we found very little information about MG effects in the ellipticity distributions.
Because of this, and because we are interested in the spherically averaged density profile of voids, we considered spherical voids to be more suitable for this analysis.
In principle, this also helps us to parameterize the void density profiles more simply.

According to spherical expansion theory, in an Einstein-de Siter (EdS) cosmology, voids are predicted to have an average overdensity $\Delta_{\text{v}} = 0.2$ that reaches as far as the void radius (see Appendix A of \cite{JenningsLiHuMNRAS13}).
Even though $\Delta_{\text{v}}$ depends on cosmology and MG parameters, we set it to its EdS value as it is commonly done in the case of the density contrast for defining halos in simulations.
Moreover, in a real application, we would not know the correct gravitational theory and cosmology to compute $\Delta_v$ a priori.

The void-finding algorithm starts by computing the dark matter density on a regular grid of size $l$.
This is done by applying the cloud-in-cell (CIC) algorithm to the dark matter particles.
The overall process consists of three stages that we describe below.

1) {\bf Initial grid spheres.} First, a sphere centered on each grid cell $j$ is grown until radius $r_j$, where the mean density of the sphere reaches the critical value $\rho(r_j) \equiv 0.2\times\overline\rho_m$, which is the expected density for spherical voids at redshift $z=0$ \citep[see][]{JenningsLiHuMNRAS13}.
Here, $\overline\rho_m$ is the mean density of the dark matter particles in the cosmological box.
We refer to these spheres as grid spheres, and some of them are refined in the next step.

2) {\bf Adaptive refinement.} The second stage improves the initial estimate of the radius and the center position of the grid spheres.
This improvement is accomplished using the particle positions instead of the density of the grid cells as was done in the first stage to compute the densities.
This step aims to maximize the size of the sphere that is refined and makes the result robust regardless of the first guess given by the first stage.

We improved upon stage one by \textit{i)} growing spheres with an average density with a critical value, not only at the current position $\vec{x}_j$ (initially the center of the cell), but also at the corners of a cube of side $d$ around $\vec{x}_j$.
\textit{ii)} When one of the corners maximizes the size of the sphere that is refined, we moved $\vec{x}_j$ to that position. Otherwise we did not update $\vec{x}_j$ , but instead reduced $d$ to half of its current value (the initial value for $d$ is the grid side $l$). 
For a given sphere $j$, we iterated this refinement process (steps \textit{i} and \textit{ii})  until $d$ reached the threshold defined as the minimum value between $0.125$~Mpc/$h$ and $1\%$ of the current sphere radius $r_j$. 

When the radius $r_j$ and position $\vec{x}_j$ of a given sphere $j$ were refined, we set $r_k=0$ for every grid sphere $k$ whose center $\vec{x}_k$ was closer than $0.9 r_k$ to $\vec{x}_j$.
This was done to avoid duplicates because under refinement, these grid spheres $k$ would converge to the same values $r_j$ and $\vec{x}_j$.
We stopped refining grid spheres when none of them had a radius larger than $2 l$.
After this stage, we had what we call the catalog of void candidates (grid spheres whose radius and position were refined).

3) {\bf Family casting.} We gathered void candidates (denoted here by C) into families (which we call voids and denote by V) by applying the linking procedure described below.
In this way, a candidate added to a family becomes a void member, while a void is identified as the collection of its members.

First, we assigned the largest void-candidate to the first family. 
Then we considered the next largest candidate C, whose core was defined as the sphere around its center with a radius that is $70\%$ of the candidate's radius.
Casting then proceeded as follows:

\begin{itemize}
\item If none of the already identified voids overlapped the core of the candidate (meaning that C is isolated enough from any already detected void), we created a new family and assigned C to it.

\item If the center of C was inside an already identified void V and no other void overlapped the core of C, we assigned the candidate C to the void V.

\item Otherwise, we removed C from the candidate catalog because either it was unclear to which of the already detected voids the candidate C belongs, or because C was not isolated enough to define a new void.
If C was not discarded, it could become the linking piece of a bilobe (dumbbell-shaped) void, which is not the type of void we are interested in.

\end{itemize}

After all the spheres in the candidate catalog were cast, roughly one-third of them were discarded.
The remaining two-thirds of the original candidates were gathered into families (which we call non-spherical voids) with an average of $\text{about three}$ spheres per void.
The largest voids have nearly one hundred spherical members, while many small voids have only one spherical member (which is a consequence of the finite number of cold dark matter (CDM) particles in the simulation).

\begin{figure}
\begin{tikzpicture}

\draw[blue] (0,1.4) circle [radius=0.8];                
\fill[blue!40!white] (0,1.4) circle [radius=0.8*0.7];   
\fill[blue] (0,1.4) circle [radius=0.03];               
\draw[blue] (0,0.4) node {$c_m$};                       

\draw[blue] (0.4,3.4) circle [radius=0.5];              
\fill[blue!40!white] (0.4,3.4) circle [radius=0.5*0.7]; 
\fill[blue] (0.4,3.4) circle [radius=0.03];             
\draw[blue] (0.3,4.1) node {$c_n$};                     

\draw[blue] (2.4,2.6) circle [radius=0.8];              
\fill[blue!40!white] (2.4,2.6) circle [radius=0.8*0.7]; 
\fill[blue] (2.4,2.6) circle [radius=0.03];             
\draw[blue] (1.7,1.9) node {$c_p$};                     

\draw[blue] (4,4.1) circle [radius=0.6];                
\fill[blue!40!white] (4,4.1) circle [radius=0.6*0.7];   
\fill[blue] (4,4.1) circle [radius=0.03];               
\draw[blue] (4,4.9) node {$c_q$};                       

\draw[blue] (4.6,2.5) circle [radius=0.7];              
\fill[blue!40!white] (4.6,2.5) circle [radius=0.7*0.7]; 
\fill[blue] (4.6,2.5) circle [radius=0.03];             
\draw[blue] (5.4,2.1) node {$c_r$};                     

\draw[blue] (2.7,-0.4) circle [radius=0.4];                
\fill[blue!40!white] (2.7,-0.4) circle [radius=0.4*0.7];   
\fill[blue] (2.7,-0.4) circle [radius=0.03];               
\draw[blue] (2.1,-0.4) node {$c_s$};                       

\draw[blue] (6.3,0.2) circle [radius=0.5];              
\fill[blue!40!white] (6.3,0.2) circle [radius=0.5*0.7]; 
\fill[blue] (6.3,0.2) circle [radius=0.03];             
\draw[blue] (6.9,-0.2) node {$c_t$};                    

\draw[blue] (2.5,0.75) circle [radius=0.7];                
\fill[blue!40!white] (2.5,0.75) circle [radius=0.7*0.7];   
\fill[blue] (2.5,0.75) circle [radius=0.03];               
\draw[blue] (1.6,0.75) node {$c_u$};                       

\draw[red] (0.2,2.4) circle [radius=1.15] node {$V_i$}; 

\draw[orange] (2.2,4) circle [radius=1.5] node {$V_j$}; 

\draw[green!60!black] (4,0.4) circle [radius=1.4] node {$V_k$}; 
\draw[green!60!black] (5.3,0.5) circle [radius=1.1];    
\fill[green!60!black] (5.3,0.5) circle [radius=0.03];
\draw[green!60!black] (3.3,1.5) circle [radius=1.];     
\fill[green!60!black] (3.3,1.5) circle [radius=0.03];
\draw[green!60!black] (6.3,0.9) circle [radius=0.8];    
\fill[green!60!black] (6.3,0.9) circle [radius=0.03];

\draw[black] (5.5,4.2) circle [radius=1.3];     
\draw[black] (6.4,3.5) circle [radius=1.1];     
\fill[black] (6.4,3.5) circle [radius=0.03];
\draw[black] (6.6,2.5) circle [radius=0.9];     
\fill[black] (6.6,2.5) circle [radius=0.03];
\draw[black] (6.,3.8) node {$V_l$};             

\end{tikzpicture}
\caption{Illustration in 2D of the assignment of void candidates to families (voids).
We show voids that have already been identified $\{\,V_i\,,\,V_j\,,\,V_k\,,\,V_l\,\}$, with both $V_k$ and $V_l$ having more than one member.
The centers of each member in a family are inside the radius of another member belonging to a previous generation of the same family (we highlight this by showing the center of the family members, except for the very first detected member of each family). 
Blue circles denote void candidates and blue shades represent their cores (which correspond to $70\%$ of the candidate's radius). 
\label{fig:family_casting}
}
\end{figure}
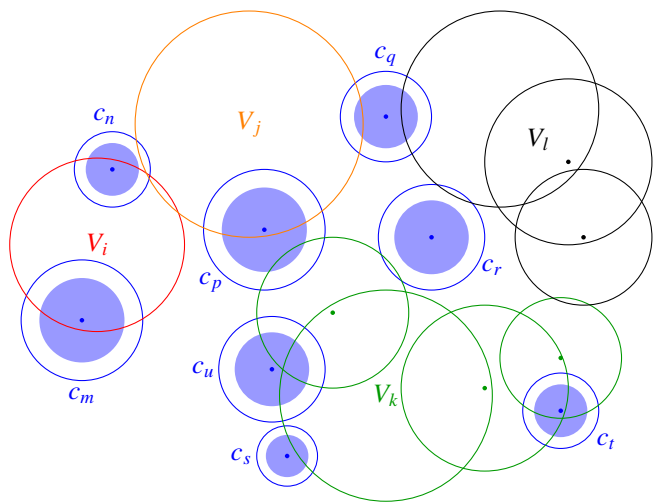

In Fig.~\ref{fig:family_casting} we illustrate the family-casting procedure.
The condition for turning a candidate C into a new void is that none of the already detected voids touches the core of C. This is the case of $c_r$, for which a new family is created, turning $c_r$ into the first member of this new family (void).
The two conditions for assigning a candidate C to an already detected family $V$ are that the center of C has to be inside $V$, and the overlap between the core of C and any other void has to be zero. This is the case for $c_m$, $c_n$ , and $c_t$.
In this case, $c_m$ and $c_n$ are assigned to family $V_i$ during the casting process, while $c_t$ is assigned to family $V_k$.
The center of $c_t$ is inside a single void (family), and no other family overlaps the blue core of $c_t$.
Finally, candidates $\{\,c_p\,,\,c_q\,,\,c_s\,,\,c_u\,\}$ are discarded during the casting process.
The centers of $c_s$ and $c_u$ are not inside any void, but they are not isolated enough from already detected voids ($V_k$ touches their cores).
On the other hand, $c_q$ is not isolated enough for becoming a new void, and its core is touched by both $V_j$ and $V_l$, therefore it is unclear to which family $c_q$ would belong. 
Likewise, the core of $c_p$ has reasonable overlap with both $V_j$ and $V_k$.

\subsection{Non-spherical and spherical voids}

The non-spherical void catalog consists of all the families that were founded by applying steps 1) to 3) described in the last subsection.
As a direct consequence of the finite number of CDM particles in the simulation, the resulting non-spherical void catalog can be seen as being composed of a subsample of small voids that are dominated by spherical members and a subsample of large voids that are dominated by non-spherical members.

We extracted a spherical-void catalog from the non-spherical catalog by defining a spherical void as the largest member of each non-spherical void.
In this case, each spherical void has a mean density equal to $0.2\times\overline\rho_m$ by construction, and its radius is denoted by $r_{0.2}$.

Extracting a spherical-void catalog from a non-spherical one instead of directly detecting spherical voids from the very beginning weakens the problem of breaking highly non-spherical voids into several spherical pieces, which causes miscounting.
Furthermore, in this case, the additional voids would inevitably be bounded by other voids, which would create a subsample with different features than those proper of voids that are bounded by overdensities.
In order to work with a purer sample, we chose to analyze the catalog of spherical voids alone.

\section{Void observables}
\label{sec:observables}

In this section we present the phenomenological expressions we used to describe the void abundance $f(\sigma)$, the void density profile $\rho(r\,,r_{0.2})$, and the void-matter linear bias $b(\sigma)$.
Here, $\sigma^2=\sigma^2(R)$ is the variance of the linear matter power spectrum smoothed on a scale $R$, which was computed for the $\Lambda$CDM and MG cases as described in \cite{RodrigoEtPRD17}.

The expressions in this section were chosen with the aim of describing the voids properties (just as the Navarro-Frenk-White halo density profile, \cite{NFW}, or some halo mass functions, see, e.g., \cite{TinkerETAJ08}, which are very useful fits for describing n-body simulations) with enough accuracy in order to constrain the MG parameters from simulated data, which is the main goal of this work.
They can lack a direct origin from first principles, but they are inspired in some theoretical results like the peak background split for describing the void-matter bias, or the excursion set formalism for the void abundance.

The specific expressions for the different void properties already depend on cosmology and gravity through the $\sigma$ function, but we found that the free parameters in these expressions depend on MG.
We take this MG dependence into account through the functional form
\begin{equation}
\tilde\gamma(x)=\left\{
\begin{array}{ll}
\gamma_1+\gamma_2\,\log_{10}(\gamma_3+x)\,, &x\equiv|f_{R0}|\text{ for $f(R)$}\,,\\
\gamma_4+\gamma_5\, \cfrac{x}{1+\gamma_6\,e^{-x^2}}\,, &x\equiv z_{SSB}\text{ for symmetron}\,,
\end{array}
\right.
\label{Eq:abundance_parameters}
\end{equation}
which are slight modifications of linear functions in $\log_{10}|f_{R0}|$ or $z_{SSB}$.
If they were pure linear functions ($\gamma_3=\gamma_6=0$), the quality of the fit for the $z_{SSB}=0$ case would be reduced, and our fiducial value for $|f_{R0}|$ would be $\sim10^{-8}$ instead of $0$.
On the other hand, $\gamma_4$ could be written as $\gamma_4=\gamma_1+\gamma_2\log_{10}(\gamma_3)$ in order to ensure a unique description for the GR limit for both $f(R)$ and symmetron, but we do not need to do this because we analyzed both theories independently.

The functional for in Eq. \ref{Eq:abundance_parameters} appears recurrently in this section.
We therefore include the MG dependence in an almost linear way because all the MG cases we study here are quite close to GR.
Mainly because no 1-to-1 map for $f(R)$ and Symmetron is available, we were unable to find a common parameterization for the MG dependence.
Furthermore, this non-equivalence between these theories is the main motivation for us to search for a probe to distinguish between them.

\subsection{Void abundance}
\label{subsec:abundance}

It has been shown \citep[e.g.,][]{RodrigoEtPRD17} that the abundance of spherical voids is well described by the excursion set formalism and parameters $\delta_{\text{v}}$ and $\delta_c$ that come from spherical expansion or collapse theories.
The voids of Voivodic et al. were grown centered on positions given by particle coordinates of the minimum of the density field given by the Voronoi volume of each particle.

We defined the center of the voids in order to maximize the void radius, therefore we obtained voids that are larger than those described in \cite{RodrigoEtPRD17}.
Moreover, because we did not set the center of our voids on the Voronoi volume particles, we detected more small voids than \cite{RodrigoEtPRD17}.
As a result, our void abundance is not well described by the excursion set model described in \cite{RodrigoEtPRD17}.
Instead, we describe the void abundance with a phenomenological formula, using the same functional form of the halo mass function as in \cite{TinkerETAJ08},

\begin{equation}
\begin{split}
\frac{dn}{d\ln R}(\sigma,x)=&\,\frac{f(\sigma,x)}{V(R)}\,\frac{d\,\ln\,\sigma^{-1}}{d\,\ln\,R}\,,\\
f(x,\sigma)=&A\,\sigma^{\tilde\gamma(x)}\,(1+\nu^b)\,e^{-c\,\sigma^{-2}}\,,\quad \nu=1.686/\sigma\,,\\
\end{split}
\label{Eq:abundance}
\end{equation}
where $\{\,A\,,\,b\,,\,c\,,\,\tilde\gamma\,\}$ need to be fit.
After the fitting process, we found that $A$, $b,$ and $c$ can be considered common constants for the theories we analyzed.
On the other hand, we needed to introduce an explicit dependency on the gravitational theory through $\tilde\gamma(x)$, Eq. \eqref{Eq:abundance_parameters}.
This allowed us to account for more large voids when MG is stronger (due to a higher level of clustering), therefore leaving less room to small voids.

On the other hand, our void-detection algorithm is not sensitive to the void-in-cloud process, which prevents small voids from surviving in dense regions \citep{JenningsLiHuMNRAS13}.
From our point of view, the void-in-cloud process is more suitable for a continuous description of the dark matter density field and is not a property of a discrete system such as an n-body simulation or a galaxy catalog.

In a discrete system, our algorithm will always find more voids when we scan over smaller scales (as long as there are enough tracers available).
This outcome is not compatible with the theoretical predictions that take the void-in-cloud process into account, but instead, it is more similar to the halo abundance functional form, Eq. \eqref{Eq:abundance}.
Furthermore, the simplest model for the void abundance corresponds to the Press-Schechter result, which has the same functional form for both voids and halos.
Therefore we expect that a more elaborate halo abundance functional form is required to describe the void number-counts in the context of our analysis.

In Fig.~\ref{fig:recovered_abundance} we show the measured void abundance and the best fit of Eq.~\eqref{Eq:abundance} for the $\Lambda$CDM, $f(R),$ and symmetron n-body simulations. 
The errors on the measurements, estimated as the variance of the octant-subsamples in the simulated box, closely follow the Poisson noise expectation of the entire sample.
We therefore used Poisson errors for the void abundance.

\subsection{Void density profile}
\label{subsec:density}

The void density profile was estimated as the mean of stacked voids traced by the dark matter particles as
\begin{equation}
\rho_v(r)=\frac{3\,m}{4\pi}\sum_i\frac{\Theta(r_i;r,\delta r)}{(r+\delta r)^ 3-(r-\delta r)^ 3}\,,
\end{equation}
with
\begin{equation}
\Theta(r_i;r,\delta r)=\left\{
\begin{array}{ll}
1\,, &\text{ when } r_i\in(r-\delta r,r+\delta r)\,,\\
0\,, &\text{ otherwise }\,,
\end{array}
\right.
\end{equation}
where $m$ and $r_i$ are the mass and position of the dark matter particles, while $2\delta r=0.05 r_{0.2}$ is the thickness of the shells we used to sample the density profile.

We split the void catalog by void sizes into seven intervals of $r_{0.2}$.
Then, we rescaled the individual density profiles from $\rho(r)\to\rho(r/r_{0.2})$.
Finally, we stacked every rescaled density profile belonging to the same size interval.
The errors on the measurements were estimated from the variance of the octants in the simulated box.

We used the following phenomenological expression \citep{profilesPRL14} to describe the void density profile: 
\begin{equation}
\frac{\rho_v(r)}{\overline\rho_m}-1=\delta_0\,\frac{1-G(y\,c)^\alpha}{1+(y\,c)^ \beta}\,,\qquad y=r/r_{0.2}\,,
\label{Eq:density_profile}
\end{equation}
where $\bar\rho_m$ is the background dark matter density, $\delta_0$ is a constant, and the parameters $\{\,\alpha\,,\,\beta\,,\,c\,,\,G\}$ depend on the void size as well as on the free parameters of the MG theory.
We parameterized them as functions of the MG parameter $x$ and $\sigma(r_{0.2})$ as follows:

\begin{equation}
\begin{split}
\alpha(x,\sigma)=\,&\tilde\alpha(x)-\alpha_0\,\sigma\,,\\
\beta(x,\sigma)=\,&\beta_0\,\alpha(x,\sigma)\,,\\
c^{\beta(x,\sigma)}(\sigma)=\,&c_0\,\sigma^{-c_1}\,,\\
G(x,\sigma)=\,&\tilde G(x)+G_0\log_{10}(\,\sigma)\,,
\end{split}
\label{Eq:density_profile_parameters}
\end{equation}
$\tilde\alpha(x)$ and $\tilde G(x)$ have the same functional form as $\tilde\gamma(x)$ in Eq.~\eqref{Eq:abundance_parameters} in the $f(R)$ case, while they are linear functions of ($x=z_{SSB}$) in the case of the symmetron theory.
All the subindexed coefficients are constants, but they have slightly different values for $f(R)$ and symmetron because we analyzed each theory independently in this work.
In Eq.~\eqref{Eq:density_profile_parameters}, $\sigma\equiv\sigma(r_{0.2})$ contains cosmological and MG information as well as the void size, while $x$ is the MG parameter itself (see Eq.~\eqref{Eq:abundance_parameters}).
As a result, the $f(R)$ (symmetron) theory has a total of $11$ ($13$) parameters to be fitted.

We fit for the free parameters in seven stacked profiles for each one of the different values of $|f_{R0}|$ or $z_{SSB}$.
The seven stacked profiles for each MG case include all the voids in the same bin.
These seven bins share the same length in $\log(r_{0.2})$, and they split the void sample $r_{0.2}\in(1.5\,Mpc/h\,,14\, Mpc/h)$ by size.
The value of the radius reported for a given stack corresponds to the mean of the void radius in each bin.
On the other hand, the density was measured for $\{150,140,130,120,110,100,90,80\}$ spherical shells of thickness $0.11/r_{0.2}$ for the smallest to the largest voids associated with the seven bins defined above.
In the next paragraphs, we describe the heuristic choices we made for the particular parametrization in Eq. \eqref{Eq:density_profile_parameters}.

$\alpha\,,\beta$) For $r\gtrsim 2\,r_{0.2}$ the expression \eqref{Eq:density_profile} falls like $r^{\alpha-\beta}$, in our case as $r^{(1-\beta_0)\alpha}$ with $(1-\beta_0)<0$ and $\alpha>0$, describing the tail of the profile.
This tail is steeper for larger voids because it is the linear void-matter correlation function for larger scales, and we approximated this effect with the $\sigma$ dependence of $\alpha$.
The MG dependence in the term $\tilde\alpha(x)$ was added because the steepness and amplitude of the matter-void correlation function is systematically higher for stronger modifications in the gravity force.

$c$) The value of $c^{-\beta}$ sets the position of the peak in expression \eqref{Eq:density_profile}.
Our voids are defined to have the same mean density (below the background level) at $r=r_{0.2}$, which causes their peaks to stand away from the center in the cases of higher walls (smaller voids or stronger MG).
This effect is taken into account by letting $c^{-\beta}$ increase with $\sigma$ (decrease with $r_{0.2}$).

$G$) Finally, the second term in $G$ allows us to move from a positive to a negative density around and beyond the wall of the void when we consider larger voids.
Again, $\tilde G(x)$ lets us mimic the effect that MG shows stronger clustering on small scales, which causes the change of sign to occur for higher values of $r_{0.2}$.

Some void density profiles are shown in Figs. \ref{fig:recovered_profile_f(R)} and \ref{fig:recovered_profile_symm}.
We show $\rho_v(r/r_{0.2}) /\bar\rho_m -1$ for three different void sizes and the different MG cases.

We set the density at the void center $\delta_0$ to a constant value because we did not take the inner part of the profile into account in this analysis. 
We computed $\delta_0$ as being the average, over all the void sizes and gravity cases, of the central density by fitting a power law in $r/r_{0.2}$ plus a constant to the inner void profiles.
This was done because Eq.~(\ref{Eq:density_profile}) cannot describe the inner and outer parts of the profiles very well simultaneously for the spherical voids because there is an abrupt change in the measured profiles around $r=r_{0.2}$ that is due mainly to the steepness of the walls.
Still, this functional form follows the shape of density profile (of the voids analyzed in this work) better than other proposals in the literature \citep{10.1093/mnras/263.2.481,Maggiore_2010,10.1111/j.1365-2966.2005.09064.x,Lavaux_2012,PhysRevLett.110.021302,10.1093/mnras/stt1069,10.1093/mnras/stu307}.

\subsection{Matter-void linear bias}
\label{subsec:bias}

The linear bias between the dark matter density field and the void distribution on large scales was estimated as
\begin{equation}\label{Eq:bias_estimator}
b = \left. \cfrac{P_{mv}(k)}{P_{mm}(k)}\right|_{k\to0}\,,
\end{equation}
where $P_{mm}$ is the matter power spectrum and $P_{mv}$ is the matter-void cross spectrum.
An alternative definition for the void bias would be $b=\sqrt{P_{vv}/P_{mm}}$.
We did not use the last expression for two reasons.
First, the estimation of $P_{vv}$ is prone to shot-noise much higher than $P_{vm}$.
Second, on large scales, $P_{vm}$ changes sign for voids with $r_{0.2} \sim 7$  Mpc$/h$, while $P_{vv}$ is always positive by definition.
This change of sign in $b$ is expected by the peak-background-split (PBS) prediction \citep{zobov-voidsRD14} and has been seen in simulations \citep{HamausETPRL14}.
Therefore, using $P_{vm}$ to estimate the void-matter bias allows us to use a wider range of void sizes, which gives better constraints.

The PBS approach (though mainly for describing halos) consists of splitting the total matter density field into two independent components: a long-wavelength background contribution, and a short-wavelength peak contribution.
In this scenario, the number of peaks will change differentially with the background contribution for a given density threshold that defines the peaks.
This difference can be translated into a linear bias between the total density field and the number density of the tracer (the peaks).

When a universal halo mass function there is available, the PBS can be used to compute a self-consistent linear halo-matter bias.
A mass function as predicted by the excursion set formalism and a PBS bias both describe roughly simulated data and provide indications for constructing more accurate phenomenological counterparts.
The same methods have been broadly applied in the literature to the void description \citep[see, e.g.,][]{ShethWeyngaetMNRAS04,zobov-voidsRD14}, which we take as a frame to set up more flexible expressions to tightly fit the n-body simulations including the MG effects.

We parameterize the linear void-matter bias as
\begin{equation}
b(r_{0.2}) = a + \tilde c(x) \,\sigma^{-2} + d \,\sigma^{-4}\,,\qquad \sigma=\sigma(r_{0.2})\,,
\label{Eq:bias}
\end{equation}
where $a$ and $d$ are constants, while $\tilde d=\tilde c(x)$ has the same functional form as $\tilde \gamma(x)$ in Eq.~\eqref{Eq:abundance_parameters} for the $f(R)$ case, but is a linear function of ($x=z_{SSB}$) for the symmetron case.

The results of the next section show that MG constraints  based on the linear bias are weaker than those from the abundance and the density profile analyses because the length of the simulation box is $256$ Mpc/$h$, giving a minimum Fourier mode $k=0.025$ $h$/Mpc and then a poor sampling of large scales.
We used the first five linear bins of length $0.05$ Mpc/$h$ to fit the linear trend of the matter-void bias, as shown in Fig.~\ref{fig:bias_lcdm} for the $\Lambda$CDM case.
The errors associated with spectra $P_{vm}$ and $P_{mm}$ were estimated from the variance of all the modes that contribute to a given bin in Fourier space, while the errors on the linear bias come from the fit of Eq.~\eqref{Eq:bias_estimator} as a linear function of $k^2$ for $k<0.25$ $h$/Mpc (see Fig.~\ref{fig:bias_lcdm}).

\clearpage

\section{Constraining modified gravity}
\label{sec:right_analysis}

\begin{figure}
        \includegraphics[width=8cm]{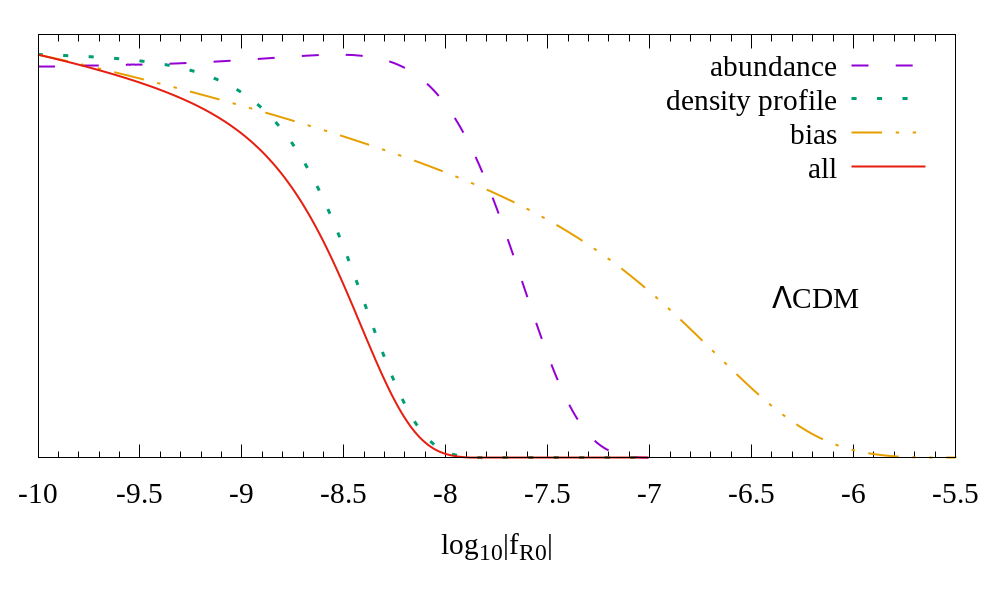}

        \includegraphics[width=8cm]{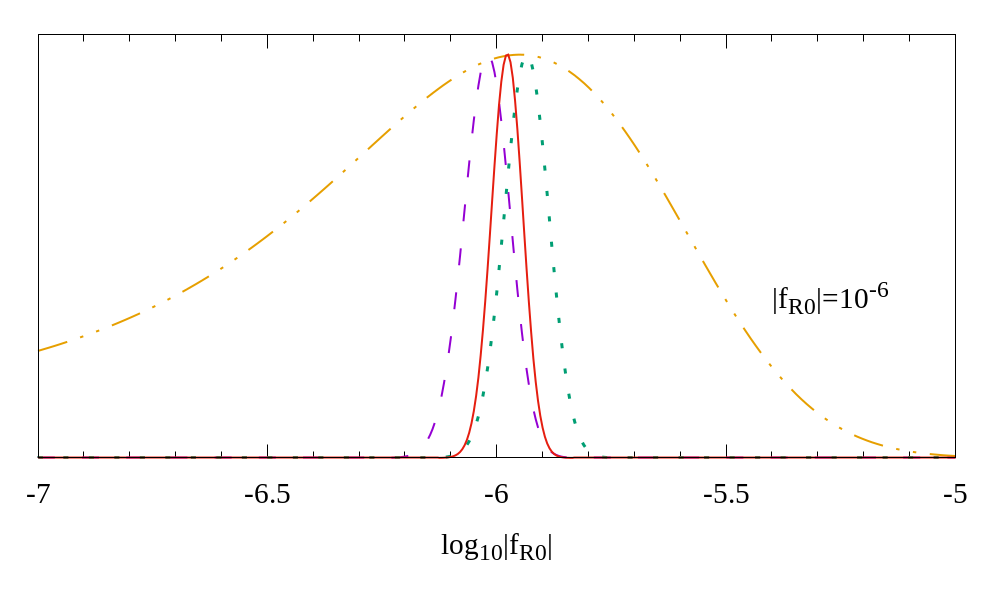}

        \includegraphics[width=8cm]{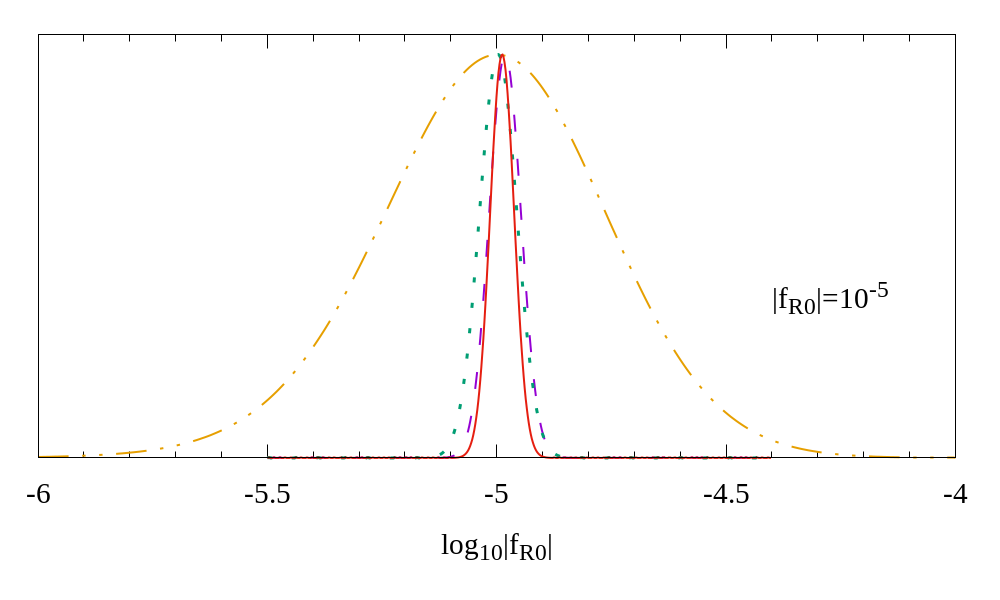}

        \includegraphics[width=8cm]{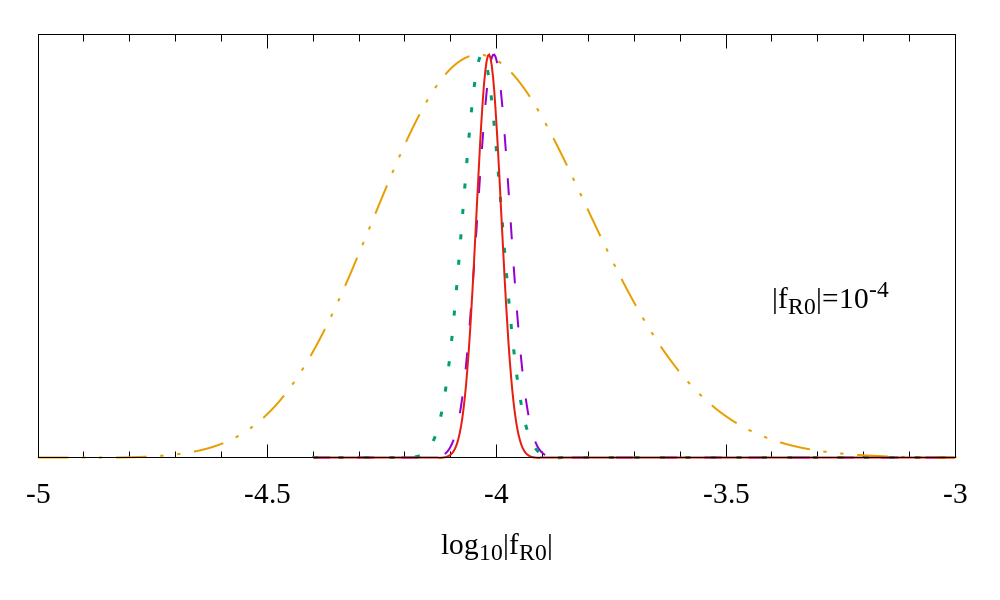}
    \caption{Posterior distributions for the free parameter of the $f(R)$ theory when we analyzed the $\{\,\Lambda\text{CDM}\,,\,f6\,,\,f5\,,\,f4\,\}$ simulations from top to bottom.
    We note that $|f_{R0}|$ constraints from abundance and density profile are compatible at $\{\,1\,,\,2\,,\,1\,,\,1\,\}$ $\sigma$ for the $\{\,\Lambda\text{CDM}\,,\,f6\,,\,f5\,,\,f4\,\}$ cases; see Table \ref{tab:tension}.
    }
    \label{fig:recover_f}
\end{figure}

\begin{figure}
        \includegraphics[width=8cm]{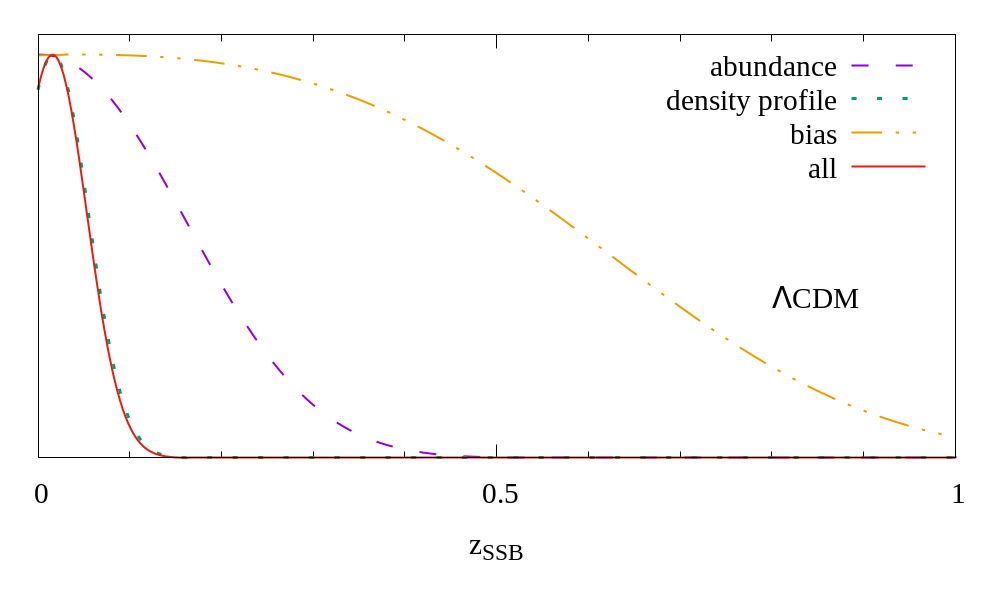}

        \includegraphics[width=8cm]{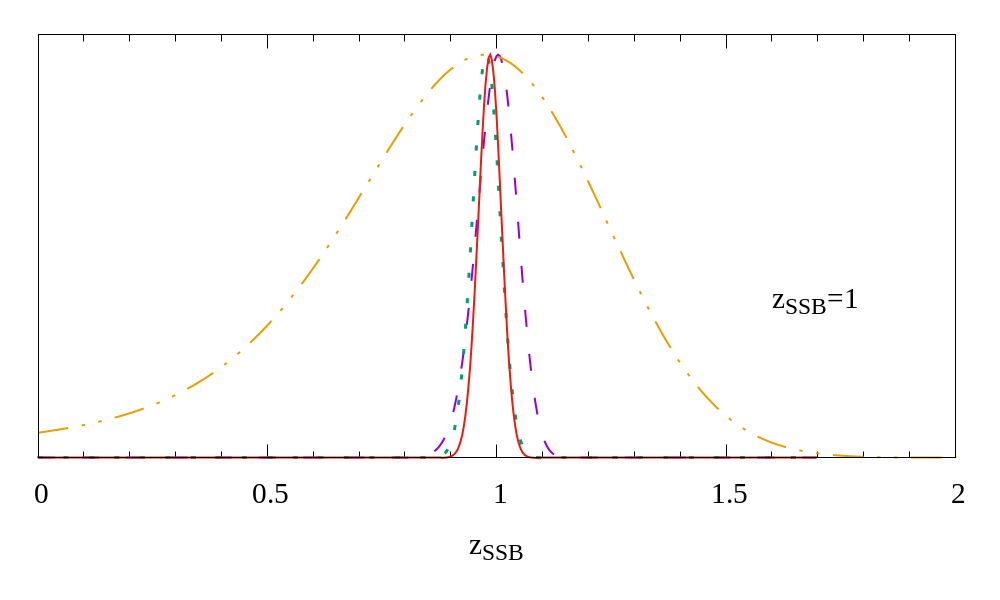}

        \includegraphics[width=8cm]{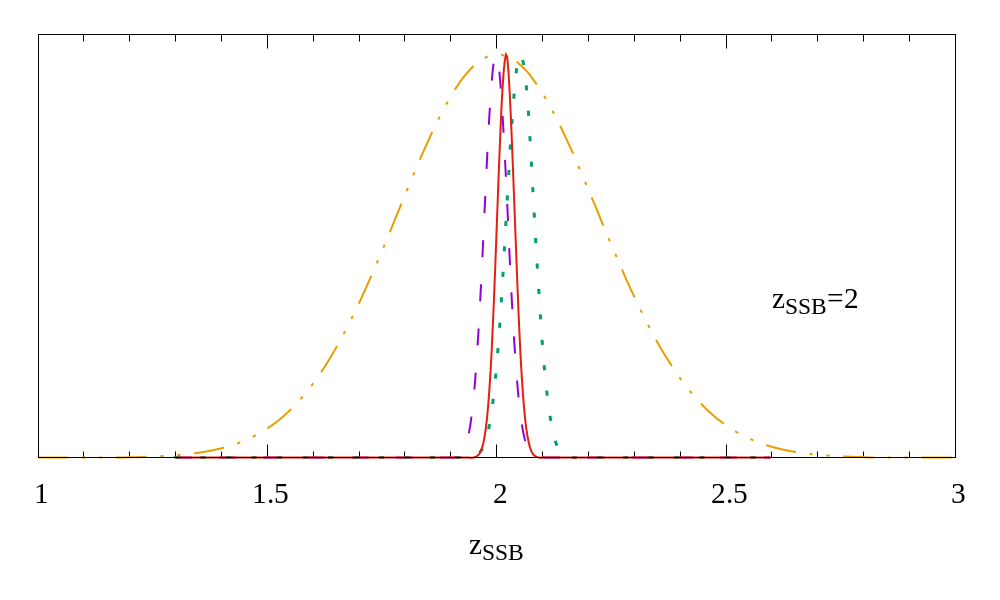}

        \includegraphics[width=8cm]{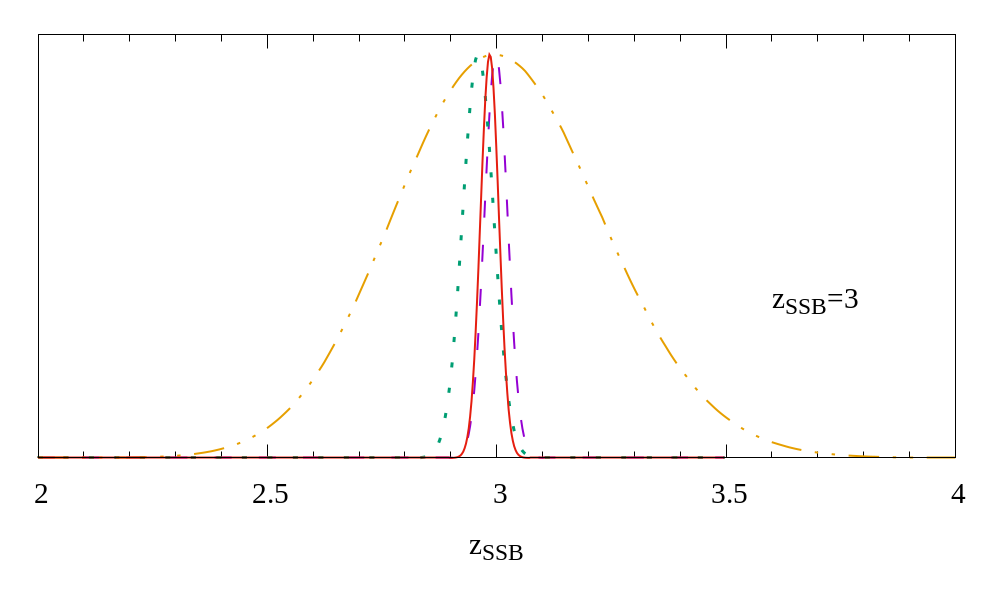}
    \caption{Posterior distributions for the free parameter of the symmetron theory.
    The $z_{SSB}$ constraints from abundance and density profile are compatible at $\{\,1\,,\,1\,,\,2\,,\,1\,\}$ $\sigma$  for the $\{\,\Lambda\text{CDM}\,,\,A\,,\,B\,,\,D\,\}$ cases, respectively, see Table \ref{tab:tension}.
    }
    \label{fig:recover_s}
\end{figure}

The first part of our analysis consisted of using the simulations to fit for the free coefficients of the phenomenological models for the void abundance Eq.~\eqref{Eq:abundance}, density profile Eq.~\eqref{Eq:density_profile}, and void-matter linear bias Eq.~\eqref{Eq:bias}, assuming the fiducial values for the MG free parameter in each case, see Tables~\ref{tab:symmetron_table} and \ref{tab:f(R)_table}.
Then, we applied the derived phenomenological models to the same simulations in order to see how well we can recover the free MG parameter in each case.
Figs.~\ref{fig:recover_f} and \ref{fig:recover_s} and Tables~\ref{tab:fl}-\ref{tab:sD} summarize the results.

\begin{figure}[!h]
\includegraphics[width=8cm]{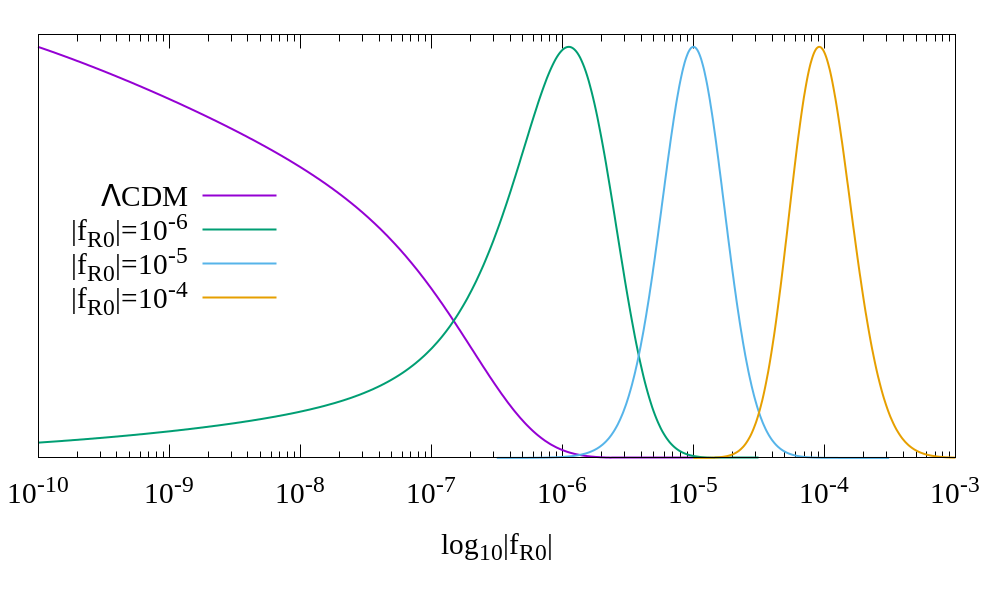}
\caption{Constraints on the $f(R)$ parameter $f_{R0}$ derived by the matter-void linear bias analysis.}
\label{fig:bias_contraints_f}
\end{figure}

\begin{figure}
        \includegraphics[width=8cm]{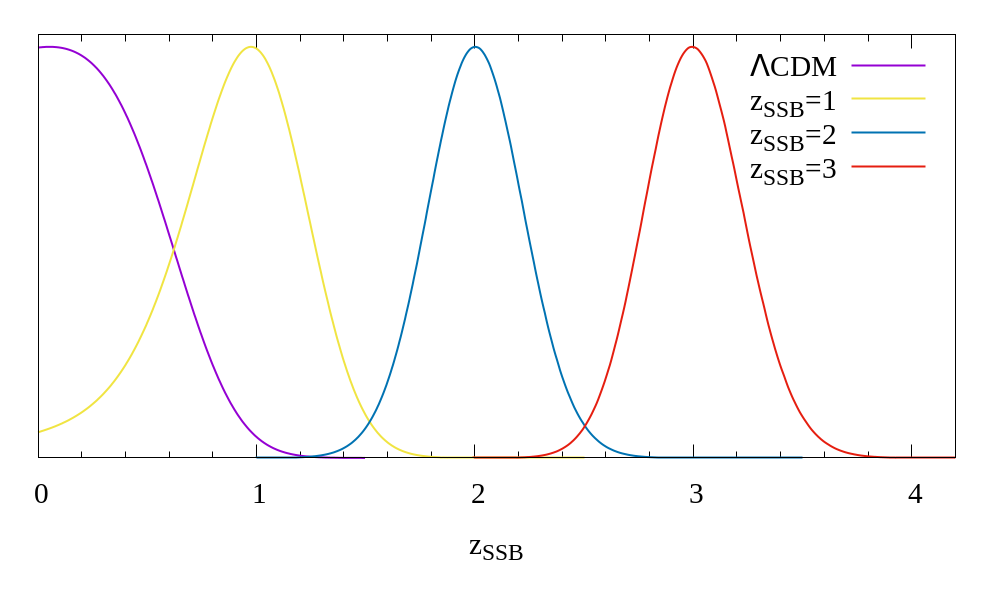}
    \caption{Constraints on the symmetron parameter $z_{SSB}$ derived by the matter-void linear bias analysis.}
    \label{fig:bias_contraints_s}
\end{figure}

We highlight that the constraints associated with the linear bias analysis are considerably weaker than those associated with the abundance or with the density profile analyses.
For this reason, we present these result in separate plots, see Figs.~\ref{fig:bias_contraints_f} and \ref{fig:bias_contraints_s}, for example.
The bias analysis shows that we can distinguish between GR and MG with parameters larger than $|f_{R0}|\gtrsim10^{-6}$ or $z_{SSB}\gtrsim1$ for the $f(R)$ or symmetron scenarios, respectively.
The results of the abundance, linear bias, and density profile analyses are compatible with each other and also with the fiducial values within two standard deviations.

In Appendix~\ref{sec:best_fits} we show the best fits of the joint analysis for the different simulations.
In these plots we show that the symmetron and the $f(R)$ effects over the three observables point in the same direction:

\begin{itemize}
\item
Stronger MG means a stronger gravitational force, which results in more clustering in intermediate scales (gravity inside halos would be the same).
Accordingly, it also means more high-mass halos because the more clustered they are, the more merging occurs.
This boost in halo clustering would translate into large voids becoming larger in MG.
As a consequence, there is less room for small voids, as we show in Fig. \ref{fig:recovered_abundance}.

\item
As was shown by \cite{RodrigoEtPRD17}, the linear power spectrum is larger in MG at small scales.
This effect translates into a matter correlation function that is also larger at intermediate and small scales in stronger MG cases.
Therefore the walls of the voids will be higher in MG\ gravity, as we show in the simulations, see Fig. \ref{fig:recovered_profile_symm}.
In consequence, because all our voids have the same mean density, the wall of the voids would also be steeper for stronger MG cases.

\item
When the linear void-matter bias is considered to depend on the void abundance, the natural consequence of having more large voids (and fewer small voids) in MG than in GR would be a higher bias.
The steeper the abundance function of a tracer, the more positive the tracer-matter bias in the PBS approximation.

\end{itemize}

These features suggest that the symmetron and $f(R)$ theories could be indistinguishable from each other when we consider only the bias, density profile, and abundance of void analyses.
To address this question, we applied the symmetron analysis to the $f(R)$ simulations and vice versa.
The results are shown in section~\ref{sec:using_wrong_model}.

\clearpage

\begin{table}
\centering
\begin{tabular}{cccccc}
\hline\hline
$\log_{10}|f_{R0}|$& best fit & mean & 1$\sigma$ & 2$\sigma$ &3$\sigma$ \\
\hline
abundance & -8.54 & -7.88 & <-7.71 & <-7.36 & <-7.14 \\ 
density profile & -10.00 & -8.68 & <-8.54 & <-8.17 & <-7.96 \\ 
bias & -10.00 & -6.80 & <-6.55 & <-6.03 & <-5.66 \\ 
all & -10.00 & -8.72 & <-8.57 & <-8.20 & <-7.98 \\

\hline
\end{tabular}
\caption{Constraints on $|f_{R0}|$ when the true value is $|f_{R0}|=0$.
For the GR simulation the $f(R)$ analysis shows that $\log_{10}|f_{R0}|<-6.03$ at the $95\%$ of confidence level when the bias analysis is applied.
}
\label{tab:fl}
\end{table}

\begin{table}
\centering
\begin{tabular}{cccccc}
\hline\hline
$\log_{10}|f_{R0}|$ & best fit & mean & 1$\sigma$ & 2$\sigma$ &3$\sigma$ \\
\hline
abundance & -6.02 & -6.02 & $\pm$ 0.05 & $\pm$ 0.10 & $\pm$ 0.15 \\ 
density profile & -5.94 & -5.94 & $\pm$ 0.04 & $\pm$ 0.09 & $\pm$ 0.14 \\ 
bias & -5.95 & -6.24 & $\pm$ 0.64 & $\pm$ 1.84 & $\pm$ 2.40 \\ 
all & -5.98 & -5.98 & $\pm$ 0.03 & $\pm$ 0.06 & $\pm$ 0.10 \\

\hline
\end{tabular}
\caption{Constraints on $|f_{R0}|$ when the true value is $|f_{R0}|=10^{-6}$.
The constraints from the abundance and density profile analyses are compatible at the $2\sigma$ level.
On the other hand, the abundance and the joint constraints are compatible with the fiducial value at $1\sigma$, while the density profile constraint is compatible at $2\sigma$.
In this case, the analysis was made as a function of $\log_{10}|f_{R0}|$.}
\label{tab:f6}
\end{table}

\begin{table}
\centering
\begin{tabular}{cccccc}
\hline\hline
$\log_{10}|f_{R0}|$ & best fit & mean & 1$\sigma$ & 2$\sigma$ &3$\sigma$ \\
\hline
abundance & -4.98 & -4.98 & $\pm$ 0.03 & $\pm$ 0.07 & $\pm$ 0.10 \\ 
density profile & -5.00 & -5.00 & $\pm$ 0.04 & $\pm$ 0.08 & $\pm$ 0.12 \\ 
bias & -5.00 & -5.01 & $\pm$ 0.24 & $\pm$ 0.50 & $\pm$ 0.80 \\ 
all & -4.99 & -4.99 & $\pm$ 0.02 & $\pm$ 0.05 & $\pm$ 0.08 \\

\hline
\end{tabular}
\caption{Constraints on $|f_{R0}|$ when the true value is $|f_{R0}|=10^{-5}$.
The constraints from the abundance and density profile analyses are compatible at the $1\sigma$ level.
Furthermore, the three individual constraints and the joint constraints are also compatible with the fiducial value at $1\sigma$.
In this case, the analysis was made as a function of $\log_{10}|f_{R0}|$.
}
\label{tab:f5}
\end{table}

\begin{table}
\centering
\begin{tabular}{cccccc}
\hline\hline
$\log_{10}|f_{R0}|$ & best fit & mean & 1$\sigma$ & 2$\sigma$ &3$\sigma$ \\
\hline
abundance & -4.01 & -4.01 & $\pm$ 0.03 & $\pm$ 0.07 & $\pm$ 0.10 \\ 
density profile & -4.03 & -4.03 & $\pm$ 0.04 & $\pm$ 0.08 & $\pm$ 0.12 \\ 
bias & -4.04 & -4.03 & $\pm$ 0.23 & $\pm$ 0.47 & $\pm$ 0.72 \\ 
all & -4.02 & -4.02 & $\pm$ 0.03 & $\pm$ 0.05 & $\pm$ 0.08 \\

\hline
\end{tabular}
\caption{Constraints on $|f_{R0}|$ when the true fiducial value is $|f_{R0}|=10^{-4}$.
The constraints from the abundance and density profile analyses are compatible at the $1\sigma$ level.
Furthermore, the three individual constraints and the joint constraints are also compatible with the fiducial value at $1\sigma$.
In this case, the analysis was made as a function of $\log_{10}|f_{R0}|$.}
\label{tab:f4}
\end{table}

\clearpage


\begin{table}
\centering
\begin{tabular}{cccccc}
\hline\hline
$z_{SSB}$ & best fit & mean & 1$\sigma$ & 2$\sigma$ &3$\sigma$ \\
\hline
abundance & 0.00 & 0.10 & <0.15 & <0.29 & <0.41 \\ 
density profile & 0.02 & 0.03 & <0.04 & <0.08 & <0.12 \\ 
bias & 0.05 & 0.32 & <0.45 & <0.80 & <1.09 \\ 
all & 0.02 & 0.03 & <0.04 & <0.08 & <0.12 \\

\hline
\end{tabular}
\caption{Constraints for $|f_{R0}|$ when the fiducial value is $z_{SSB}=0$.
In the case of the GR simulation, the symmetron analysis shows that $z_{SSB}<0.8$ at the $95\%$ of confidence level when the bias analysis is applied.}
\label{tab:sl}
\end{table}

\begin{table}
\centering
\begin{tabular}{cccccc}
\hline\hline
$z_{SSB}$ & best fit & mean & 1$\sigma$ & 2$\sigma$ &3$\sigma$ \\
\hline
abundance & 1.00 & 1.00 & $\pm$ 0.04 & $\pm$ 0.08 & $\pm$ 0.13 \\ 
density profile & 0.98 & 0.98 & $\pm$ 0.03 & $\pm$ 0.06 & $\pm$ 0.09 \\ 
bias & 0.98 & 0.93 & $\pm$ 0.29 & $\pm$ 0.62 & $\pm$ 0.77 \\ 
all & 0.99 & 0.99 & $\pm$ 0.02 & $\pm$ 0.05 & $\pm$ 0.07 \\

\hline
\end{tabular}
\caption{Constraints for $z_{SSB}$ when the fiducial value is $z_{SSB}=1$.
The constraints from the abundance and density profile analyses are compatible at $1\sigma$.
Furthermore, the three individual and the jointed constraints are compatible with the fiducial value at $1\sigma$.}
\label{tab:sA}
\end{table}

\begin{table}
\centering
\begin{tabular}{cccccc}
\hline\hline
$z_{SSB}$ & best fit & mean & 1$\sigma$ & 2$\sigma$ &3$\sigma$ \\
\hline
abundance & 2.00 & 2.00 & $\pm$ 0.02 & $\pm$ 0.05 & $\pm$ 0.07 \\ 
density profile & 2.05 & 2.05 & $\pm$ 0.03 & $\pm$ 0.06 & $\pm$ 0.09 \\ 
bias & 2.00 & 2.01 & $\pm$ 0.22 & $\pm$ 0.44 & $\pm$ 0.67 \\ 
all & 2.02 & 2.02 & $\pm$ 0.02 & $\pm$ 0.03 & $\pm$ 0.06 \\

\hline
\end{tabular}
\caption{Constraints for $z_{SSB}$ when the fiducial value is $z_{SSB}=2$.
In this case, the constraints from the abundance and density profile analyses are compatible at $2\sigma$.
On the other hand, the abundance and the jointed constraints are compatible with the fiducial value at $1\sigma$, while the density profile constraint is compatible up to $2\sigma$.}
\label{tab:sB}
\end{table}

\begin{table}
\centering
\begin{tabular}{cccccc}
\hline\hline
$z_{SSB}$ & best fit & mean & 1$\sigma$ & 2$\sigma$ &3$\sigma$ \\
\hline
abundance & 3.00 & 3.00 & $\pm$ 0.02 & $\pm$ 0.05 & $\pm$ 0.07 \\ 
density profile & 2.96 & 2.96 & $\pm$ 0.03 & $\pm$ 0.06 & $\pm$ 0.10 \\ 
bias & 3.00 & 3.00 & $\pm$ 0.22 & $\pm$ 0.45 & $\pm$ 0.68 \\ 
all & 2.98 & 2.98 & $\pm$ 0.02 & $\pm$ 0.04 & $\pm$ 0.06 \\

\hline
\end{tabular}
\caption{Constraints for $z_{SSB}$ when the fiducial value is $z_{SSB}=3$.
As in the $z_{SSB}=1$ case, the constraints from the abundance and density profile analyses are compatible at $2\sigma$.
On the other hand, the abundance and the jointed constraints are compatible with the fiducial value at $1\sigma$, while the density profile constraint is compatible up to $2\sigma$.}
\label{tab:sD}
\end{table}

\clearpage

\section{Distinguishing among gravity theories}
\label{sec:using_wrong_model}

In order to assess the level to which the three gravity scenarios GR, $f(R),$ and symmetron can be distinguished, we applied the $f(R)$ analysis to the symmetron simulations and vice versa.
This can show how well MG constraints derived from a correct model compare to those from an incorrect theory choice.

The results are summarized in Figs.~\ref{fig:recover_fs} and \ref{fig:recover_sf}.
The constraints from the abundance, density profile, and bias probes are less consistent with each other when we perform the analysis using the incorrect MG model (see Table~\ref{tab:tension}).
Similarly, as shown in Table~\ref{tab:chi2}, the values of $\chi^2/$dof are higher when we assume an incorrect model.

\subsection{Analyzing symmetron simulations using $f(R)$ theory}

In the case of a joint analysis using all void probes, the lowest $\chi^2$/dof are $\{\,1.10\,,\,1.78\,,\,2.53\,\}$ times higher for the $\{\,A\,,B\,,\,D\,\}$ simulations when we apply the $f(R)$ analysis than when the correct theory is used (see Table~\ref{tab:chi2}).
Additionally, the best-fit values for $|f_{R0}|$ from the abundance and density profile analyses disagree with each other at more than  $\{\,5\,,\,8\,,\,4\,\}$ standard deviations for the $\{\,A\,,B\,,\,D\,\}$ cases (see Table~\ref{tab:tension}).
The same result is shown in Fig.~\ref{fig:recover_fs}.

\subsection{Analyzing $f(R)$ simulations using symmetron theory}

In this case, the lowest $\chi^2$/dof from the joint analysis is $\{\,1.08\,,\,1.31\,,\,1.89\,\}$ times higher for the $\{\,f6\,,f5\,,\,f4\,\}$ simulations when they are analyzed with the symmetron theory than when $f(R)$ is used (see Table~\ref{tab:chi2}).
Likewise, the mean best-fit values for $z_{SSB}$ from the abundance and density profile analysis disagree with each other at more than $\{\,6\,,\,9\,,\,13\,\}$ standard deviations for the $\{\,f6\,,f5\,,\,f4\,\}$  cases, as shown in Table~\ref{tab:tension} and Fig.~\ref{fig:recover_sf}.

\subsection{Discussion}

We recall that when the simulations are analyzed with the correct theory, the abundance and density profile analyses agree within $\text{two}$ standard deviations, as discussed in section~\ref{sec:right_analysis}.
The results involving $\chi^2/$dof from the previous subsections show that we cannot reasonably distinguish between symmetron and $f(R)$ modified theories for the weaker MG cases we considered (i.e., $z_{SSB}=1$ and $|f_{R0}|=10^{-6}$).
The values of $\chi^2$/dof are only marginally higher when an incorrect MG model is used. 

One interesting point is that applying the correct model provides more consistency between the density profile and the abundance tests, however.
This may be due to the fact that different screening mechanisms affect the void abundance and the profile differently.
Therefore, the modeling from an incorrect MG model may effectively describe the simulated data for each void probe, but leads to conflicting values for the best fits.
In a real data analysis, this difference might indicate that an incorrect gravity model is used.

Differences might also point to observational systematics that affect multiple observables differently.
In a real analysis, we detect voids from discrete galaxies, not the dark matter field.
Therefore we expect real void catalogs to suffer from a number of observational effects that make their use for cosmological purposes much harder than what we infer here.
These effects can be collectively cast in the so-called void selection function, described by completeness and purity functions, for example, which are specific to each void-finder algorithm and survey data.
See \cite{AguenaLimaPRD18} for a discussion of how imperfect knowledge of the cluster selection function, for instance, affects cosmological constraints derived from galaxy clusters.
We expect similar effects for voids.

The key question is how well we can simultaneously constrain the parameters of the void selection function along with cosmological parameters of interest.
A particular void finder may fail to detect true voids of a given void size (lowering completeness) and may also produce false voids (lowering purity).
Misidentifications are more likely to occur for small voids or voids whose walls contain halos with a small number of galaxies. For sufficiently low-mass halos, the average number of galaxies per halo becomes $\sim1$ or lower.
In this limit, regions devoid of galaxies are not necessarily empty of dark matter.
As a result, there is no 1:1 correspondence between dark matter voids and galaxy voids.
In order to account for these effects in simulations, we need to populate dark matter simulations with galaxies and replicate all observational features such as survey mask, flux, or magnitude limit cuts.
This is beyond the scope of this work, but is necessary for accurate cosmological constraints from realistic void surveys.

For the cases where MG effects are stronger ($|f_{R0}|=10^{-4}$ and $z_{SSB}=3$), applying the correct MG theory analysis provides much better consistency between the different tests, and the minimum $\chi^2/$dof is considerably lower (compared to the incorrect MG theory). 
Therefore at the level of n-boby simulations, we can safely distinguish between symmetron cases with $z_{SSB}\gtrsim 2$ and $f(R)$ cases with $|f_{R0}|\gtrsim10^{-5}$.

\begin{figure}
        \includegraphics[width=8cm]{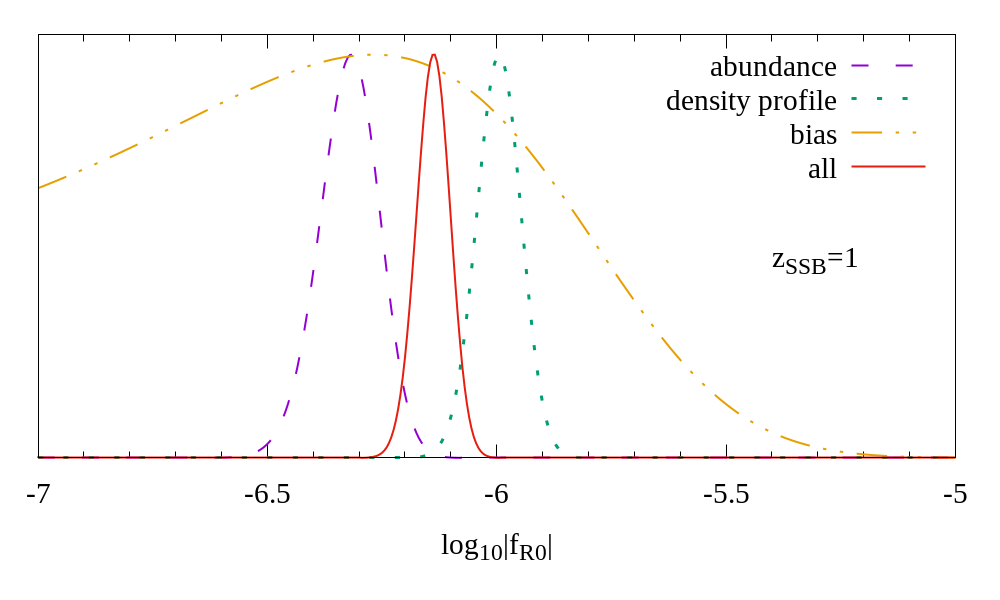}

        \includegraphics[width=8cm]{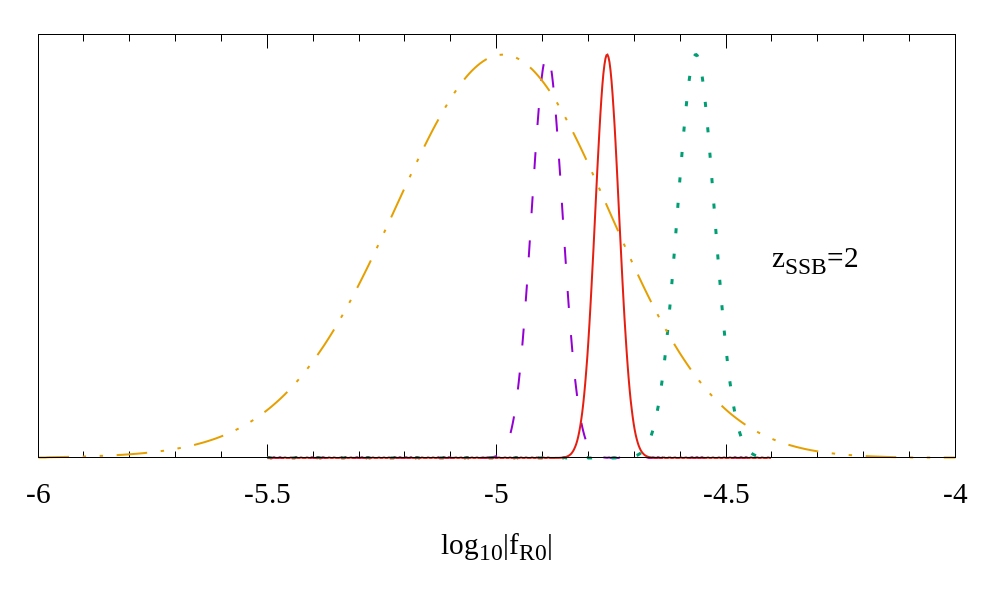}

        \includegraphics[width=8cm]{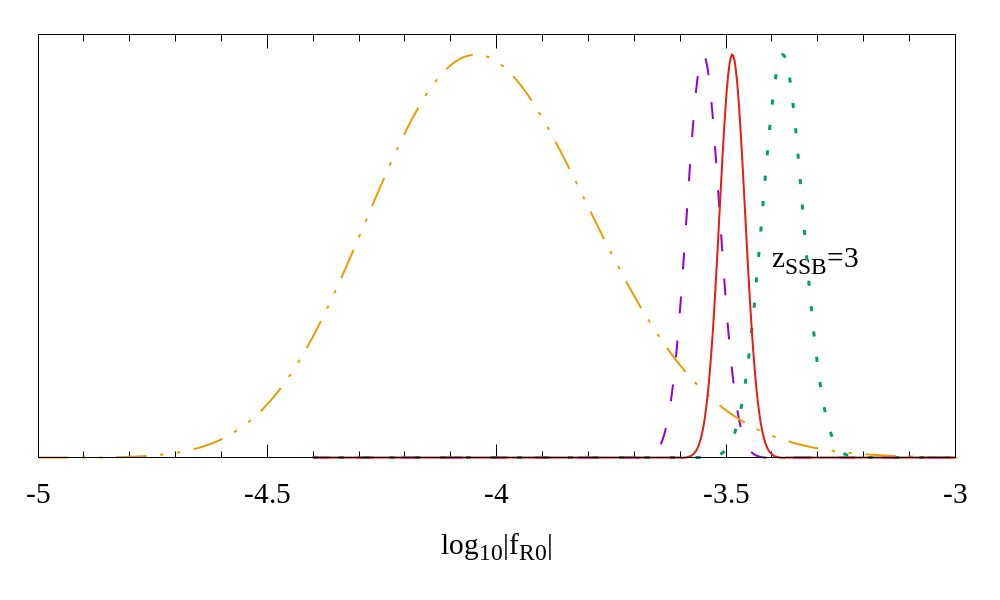}
    \caption{Recovered value for $|f_{R0}|$ for the symmetron simulations.
    }
    \label{fig:recover_fs}
\end{figure}

\begin{figure}
        \includegraphics[width=8cm]{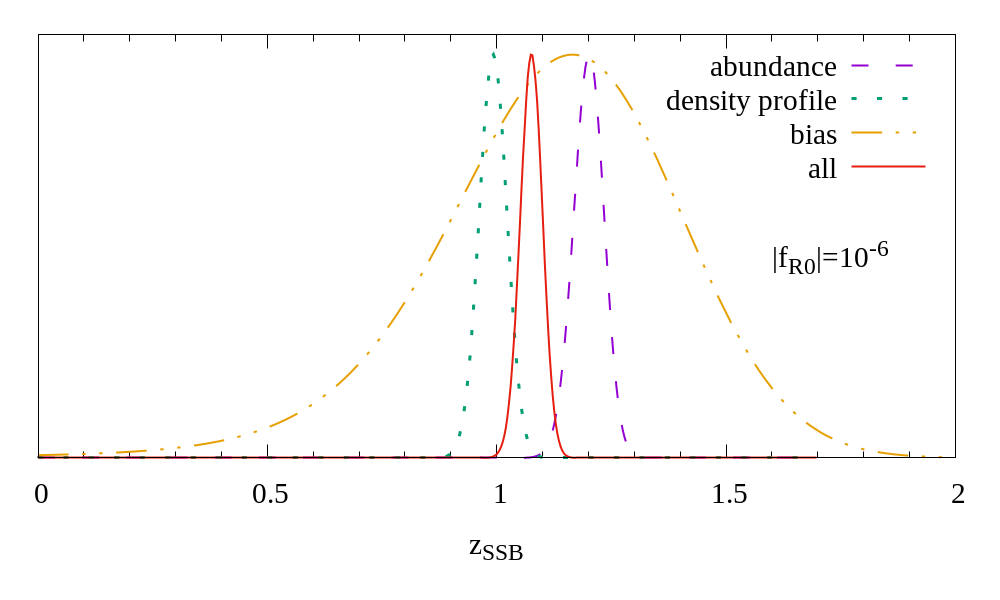}

        \includegraphics[width=8cm]{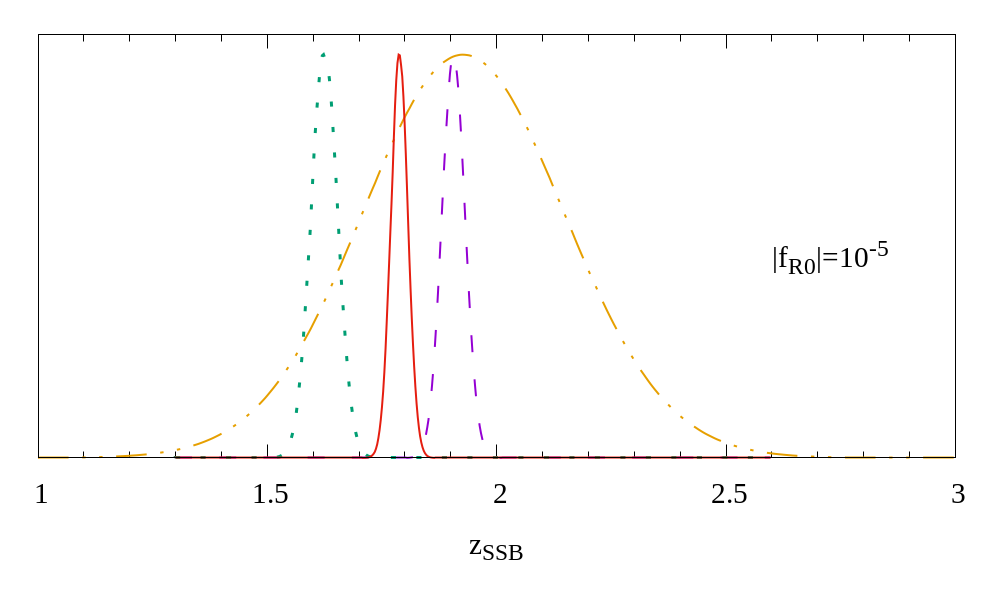}

        \includegraphics[width=8cm]{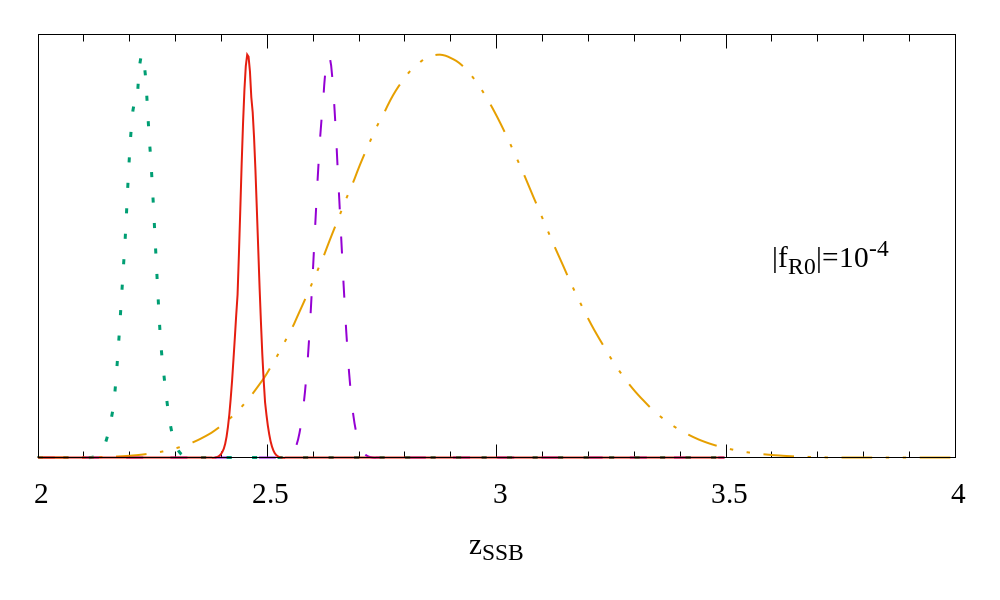}
    \caption{Recovered value for $z_{SSB}$ for the $f(R)$ simulations.}
    \label{fig:recover_sf}
\end{figure}

Interestingly, Figs.~\ref{fig:bias_contraints_s} - \ref{fig:recover_sf} show that any of the three void probes can tell us whether gravity is modified or not because none of the MG cases is compatible with $\Lambda$CDM, even when an incorrect MG theory is assumed.
Moreover, when we analyze the $\Lambda$CDM case, we recover $|f_{R0}| = 0$ or $z_{SSB} = 0$.
This means we will be able to know whether the Universe is ruled by GR or MG after applying any of the void analyses we considered here, even though it may be challenging to distinguish between the weakest cases of $f(R)$ and symmetron by the joint analysis of the linear bias, density profile, and abundance of voids alone.
On the other hand, if MG is stronger than the weakest cases considered here, the individual analyses may indicate differences in their best fits, forcing us to consider other MG models.

Finally, we note that the degeneracy between the weakest cases of MG analyzed here might in principle be broken by combining our void analysis with information from halo properties, such as their abundance, bias, and profiles.
Recently, the splashback radius \citep{Adhikari18, Contigiani18} and the turnaround radius \citep{Turnaround-Virial18, Turnaround-f(R)18} of halos have been shown to be signficantly affected by MG effects.

\begin{table}[!h]
\centering
\begin{tabular}{ccc}
\hline\hline
 & Analyzed with the   & Analyzed with the\\
 & $f(R)$ model & symmetron model \\
 \hline
Case & $\chi^2$/dof & $\chi^2$/dof \\
\hline
$|f_{R0}|=10^{-4}$ & \textbf{0.66} & 1.25 \\ 
$|f_{R0}|=10^{-5}$ & \textbf{0.50} & 0.66 \\ 
$|f_{R0}|=10^{-6}$ & \textbf{0.53} & 0.57 \\ 
$\Lambda$CDM & 0.52 & 0.58 \\ 
$z_{SSB}=1$ & 0.63 & \textbf{0.57} \\ 
$z_{SSB}=2$ & 0.94 & \textbf{0.53} \\ 
$z_{SSB}=3$ & 1.45 & \textbf{0.57} \\ 

\hline
\end{tabular}
\caption{$\chi^2$ per degree of freedom for the joint analysis using all void observables.
We highlight in bold the results of the cases analyzed with the correct MG theory.
}
\label{tab:chi2}
\end{table}

\begin{table}[!h]
\centering
\begin{tabular}{ccc}
\hline\hline
 & Analyzed with   & Analyzed with\\
 & $f(R)$ model & Symmetron model \\
 \hline
Case & $K$ & $K$ \\
\hline
$|f_{R0}|=10^{-4}$ & \textbf{0.70} & 13.50 \\ 
$|f_{R0}|=10^{-5}$ & \textbf{0.39} & 9.45 \\ 
$|f_{R0}|=10^{-6}$ & \textbf{1.70} & 6.91 \\ 
$\Lambda$CDM & 0.17 & 0.09 \\ 
$z_{SSB}=1$ & 5.33 & \textbf{0.68} \\ 
$z_{SSB}=2$ & 8.14 & \textbf{1.73} \\ 
$z_{SSB}=3$ & 4.27 & \textbf{1.30} \\ 

\hline
\end{tabular}
\caption{$K=|\bar x_{density}-\bar x_{abundance}|/\text{max}\{\,\sigma_{density}\,,\,\sigma_{abundance}\,\}$ indicates the level difference in the MG parameter $\bar x$ recovered by the density and the abundance analyses with uncertainty $\sigma$.
Again, we highlight in bold the results of the cases analyzed with the correct MG theory.}
\label{tab:tension}
\end{table}

\clearpage

\section{Conclusions}
\label{sec:conclusions}

We used a spherical-void finding algorithm to construct void catalogs in n-body simulations of GR as well as $f(R)$ and symmetron theories.
We measured the abundance, bias, and profiles of these voids and modeled our measurements with phenomenological fitting formulae.
We then used these expressions and the simulated data to assess how well MG models can be constrained from void properties.

The void-finding algorithm we used can be described by the following main points:
First, we searched for underdense spheres with a mean density given by $\overline\rho=0.2\overline\rho_m$ (which defines the void radius $r_{0.2}$), centered on the cubic pixels of a regular grid.
The pixel size was set here to be half of the mean particle distance, therefore we can take advantage of the full resolution of the simulations.
Next, we maximized the radius of each sphere by refining the position of its center.
Non-spherical voids were defined as the union of spheres with a neighbor-to-neighbor overlap larger than a given threshold, while spherical voids were selected as the largest sphere of the non-spherical ones.

The voids we found show an abrupt change in the density profile around the void radius.
This makes it difficult to fit both the inner and outer regions of the void profile simultaneously.
On the other hand, the height of the void walls increases in MG scenarios, and its variation with MG parameters is much more significant than the inner profile variation.
Therefore, we chose to use only the outer part of the profile in our analysis.

Clearly, the changes due to MG observed in void properties are connected to those observed in halo properties. 
It is well known that the matter power spectrum and the halo properties change quite significantly as a function of MG parameters \citep[e.g.,][]{SchmidtETPRD09, WymanETPRD13}.
Viable MG models increase gravity effects, causing massive halos to become more abundant and voids to become emptier.
Because the most underdense regions ($\delta \sim -1$) in a GR scenario cannot be much emptier in MG, the inner regions of voids do not change significantly, and most of the void profile modifications are concentrated around the void walls.
Likewise, void radii are larger in MG than in GR, which is also compatible with the MG effects on halo properties.
The more massive halos are more clustered on void walls in MG, producing higher walls and larger voids.

We parameterized the void abundance, density profile, and linear bias as functions of the linear power spectrum rms $\sigma(r_{0.2})$ and the MG free parameter (either $f_{R0}$ or $z_{SSB}$).
We fit for the free coefficients of these parameterizations using the measurements made on sets of n-body simulations.
%
%
Applying these parameterizations to analyze the same simulations, we recovered values for the MG free parameters that are compatible with the true values within 2$\sigma$ in the case of a joint analysis including all void probes (abundance, density profile, and linear bias).
Additionally, the values of the MG parameter coming independently from void abundance and void density profile analyses are also compatible with each other within two standard deviations.


The constraints on MG parameters from the linear bias are weaker than those from the density profile and abundance analyses, mainly because we analyzed relatively small box simulations, that is, cubic boxes with side $256$ Mpc$/h$.  This provides a poor sampling of Fourier modes on linear scales.
Nevertheless, the constraints on MG parameters from the bias analysis show that we can distinguish between GR and a $f(R)$ model with $|f_{R0}|> 9.3\times10^{-7}$ at the $95\%$ confidence level.
Similarly, we can distinguish between GR and a symmetron model with $z_{SSB}>0.8$ at the $95\%$ confidence level.


We also applied the symmetron analysis to the $f(R)$ simulations and vice versa.
For the MG scenarios closest to GR, that is, $z_{SSB}=1$ and $|f_{R0}|=10^{-6}$, we were unable to significantly distinguish between symmetron and $f(R)$  using any of the void properties we analyzed, even though we can distinguish them from GR.
However, analyzing the simulations with an incorrect theory causes a difference in the MG parameter best fits inferred from individual probes, indicating that the MG model used is inappropriate.

For the other MG scenarios,  $z_{SSB}=\{\,2\,,\,3\,\}$ and $|f_{R0}|=\{\,10^{-5}\,,\,10^{-4}\,\}$, we can distinguish among $f(R)$, symmetron, and GR based mainly on two features.
First, the MG parameters posterior distributions for the abundance and the density profile are compatible with each other within two standard deviations when the correct model is used, but with an incorrect theory, they are inconsistent by over four standard deviations.
Second, the minimum $\chi^2$/dof is between $1.3$ and $2.5$ times larger when an incorrect theory is applied.

Finally, the joint analysis shows a difference of over three standard deviations between GR and the weakest modification on the MG models we analyzed.
This type of analysis appears a promising tool for distinguishing gravity models, but further studies must be made, including realistic observational conditions.
We expect the combination of void and halo properties to be particularly useful for constraining and distinguishing MG models. 
Because halos and voids respond differently to the increased forces and the screening effects that are unique to each model, the joint analysis of halo and void properties is expected to provide important consistency tests and help break degeneracies in parameter space.
We hope to address some of these questions in future work.

\begin{acknowledgements}
We thank Claudio Llinares for providing the n-body simulations used in this work.
This work has made use of the computing facilities of the Laboratory of Astroinformatics (IAG/USP, NAT/Unicsul), whose purchase was made possible by the Brazilian agency FAPESP (grant 2009/54006-4) and the INCT-A.
EP and RV are supported by FAPESP. 
ML is partially supported by FAPESP and CNPq.
DFM acknowledges support from the Research Council of Norway, and the NOTUR facilities. 
\end{acknowledgements}



\bibliographystyle{aa}
\bibliography{ref}


\begin{appendix}
\section{Best fits}
\label{sec:best_fits}

In this appendix we show the abundance, density profile, and bias associated with the best-fit MG parameters (see Figs.~\ref{fig:recover_f} and \ref{fig:recover_s}) recovered from the joint analysis of the three void properties.


\begin{figure}
        \includegraphics[width=8cm]{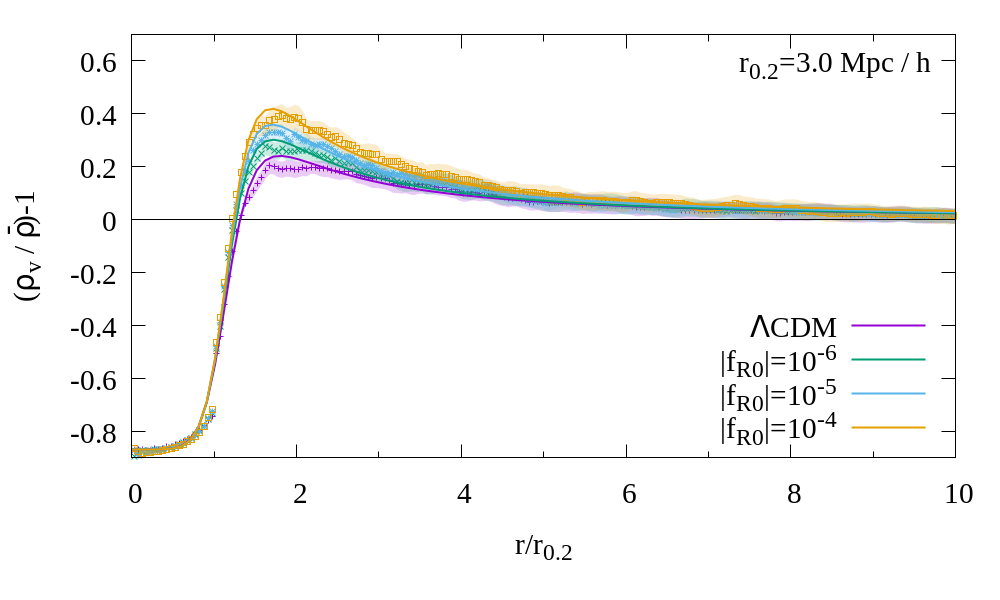}

        \includegraphics[width=8cm]{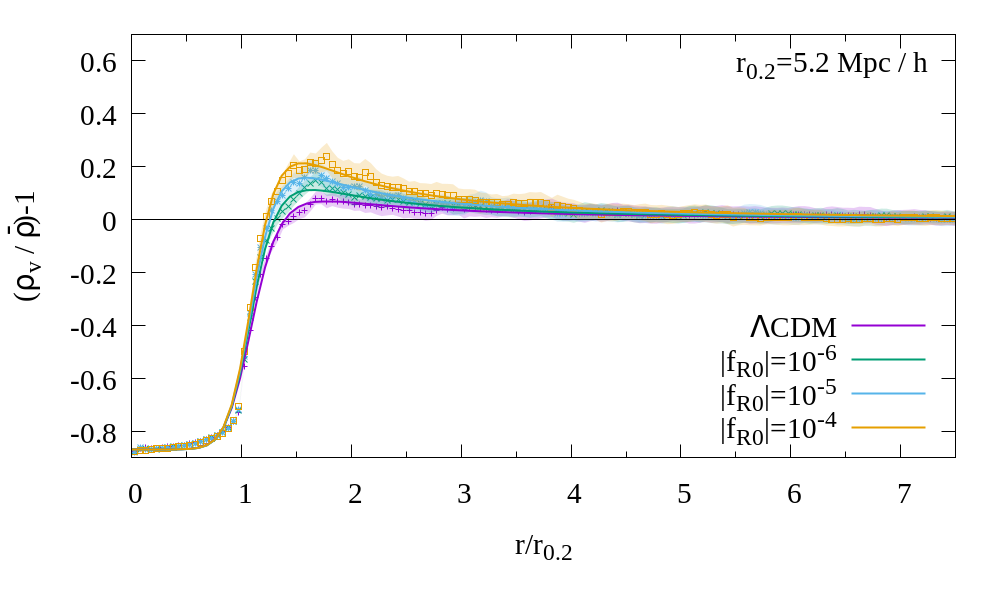}

        \includegraphics[width=8cm]{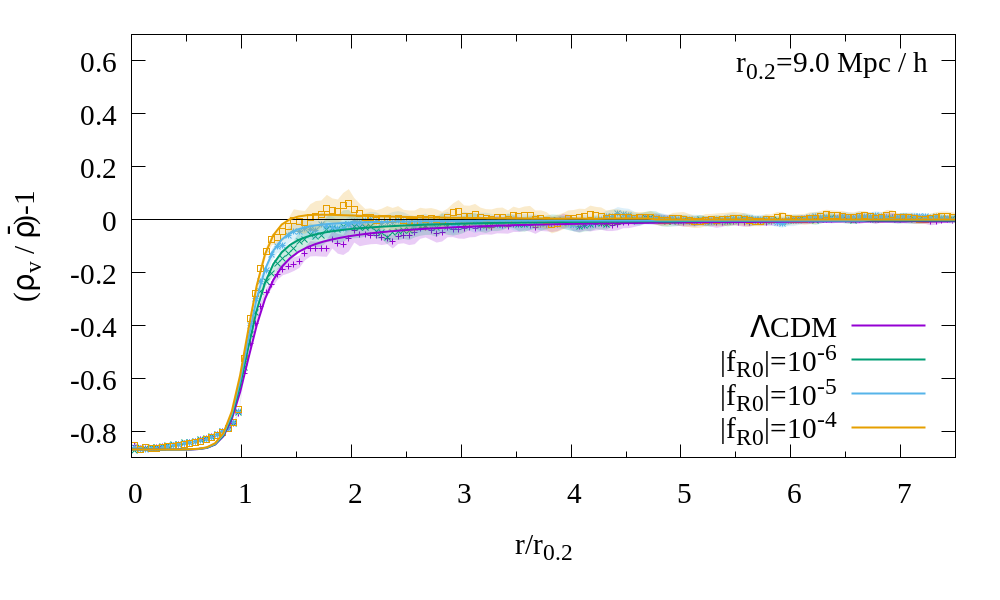}
    \caption{
    Void density profiles measured in the $f(R)$ simulations (points), and best fit of the model in Eq. \eqref{Eq:density_profile} (lines).
    We split the void catalog into seven subsamples, corresponding to different void sizes, then we stacked all the void density-profiles for each subsample. 
    Here we show the stacked subsamples with mean radius $r_{0.2}=3.0$ Mpc$/h$ (top panel), $5.2$ Mpc$/h$ (middle panel), and $9.0$ Mpc$/h$ (bottom panel).
    }
    \label{fig:recovered_profile_f(R)}
\end{figure}

\begin{figure}
        \includegraphics[width=8cm]{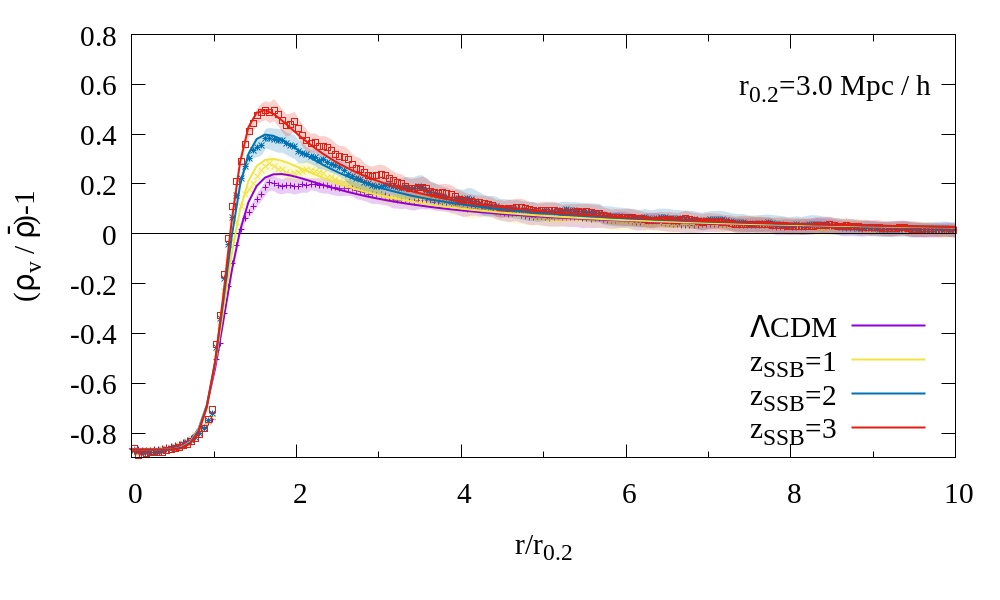}

        \includegraphics[width=8cm]{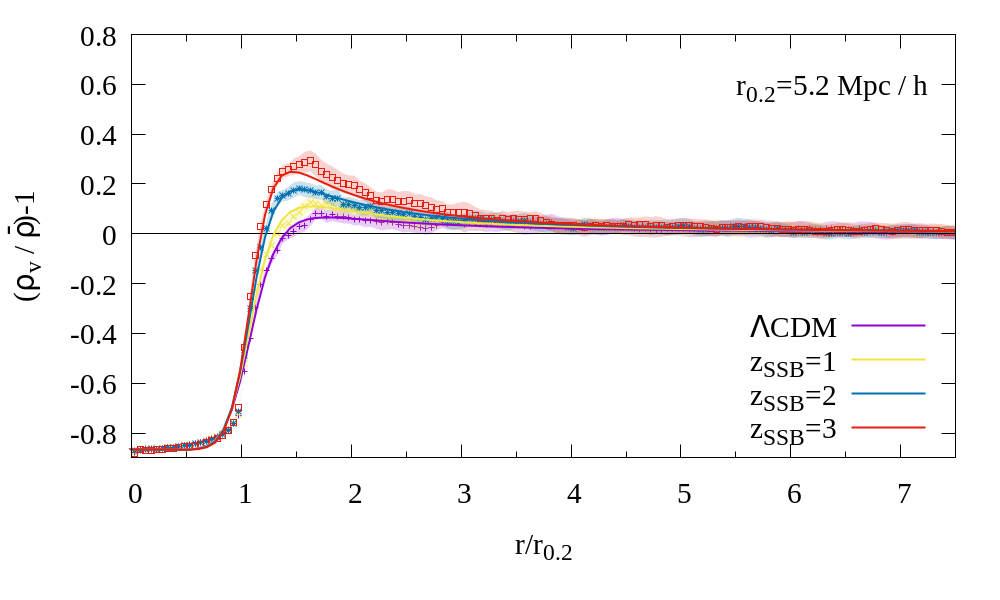}

        \includegraphics[width=8cm]{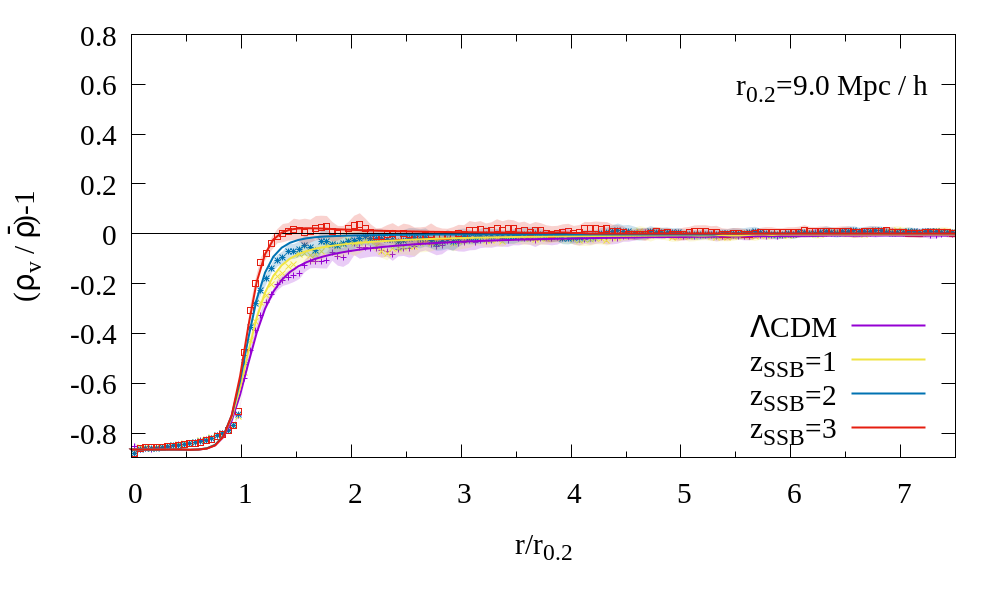}
    \caption{Same as Fig.~\ref{fig:recovered_profile_f(R)}, but for the symmetron simulations. In every case, the void wall is higher for stronger deviations of the modified gravity with respect to GR.}
    \label{fig:recovered_profile_symm}
\end{figure}


\begin{figure}
        \includegraphics[width=8cm]{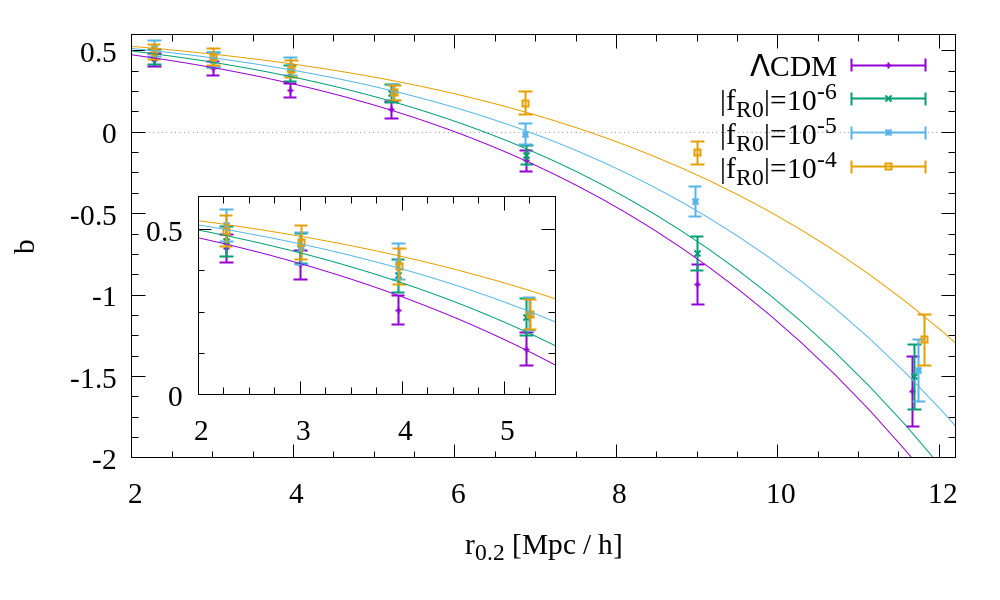}

        \includegraphics[width=8cm]{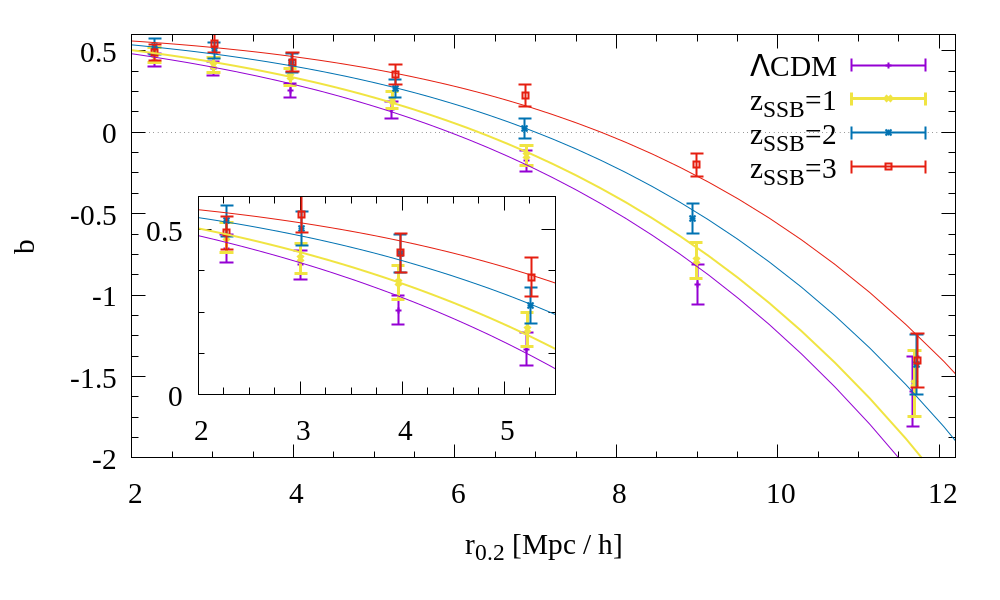}
    \caption{Linear matter-void bias for the $f(R)$ and the symmetron cases.
    The points and error bars represent the measurements from simulations described in section \ref{subsec:bias}, while the solid lines represent the best fit of $|f_{R0}|$ or $z_{SSB}$ in the $f(R)$ or symmetron cases, respectively.}
    \label{fig:bias}
\end{figure}

\begin{figure}
        \includegraphics[width=8cm]{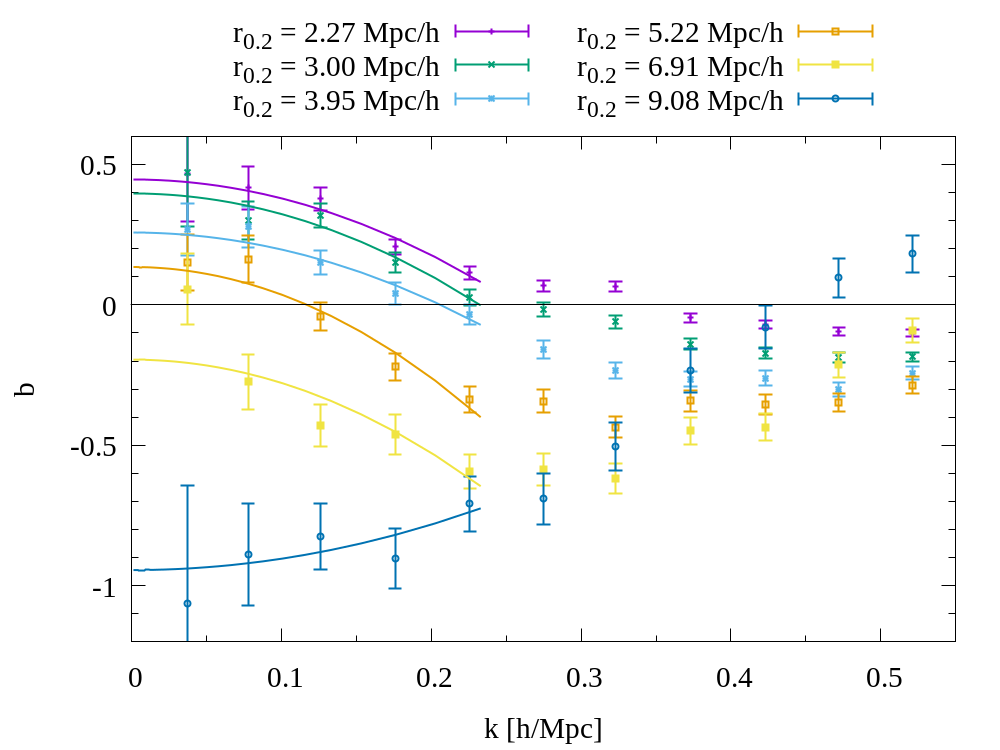}
    \caption{Matter-void bias as a function of scale $k$ for voids with different sizes given by the value of $r_{0.2}$ in the $\Lambda$CDM simulation. Points denote simulation measurements and lines represent the best fit of the large-scale trend. A linear function in $k^2$ was fit for large-scale modes with $k<0.25$ $h$/Mpc.}
    \label{fig:bias_lcdm}
\end{figure}


\begin{figure}
        \includegraphics[width=8cm]{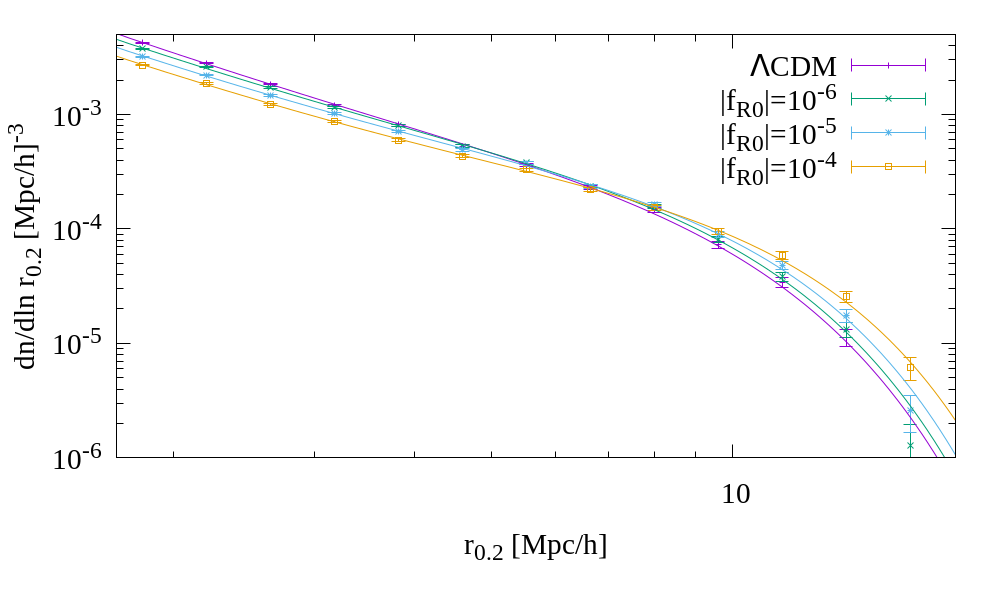}

        \includegraphics[width=8cm]{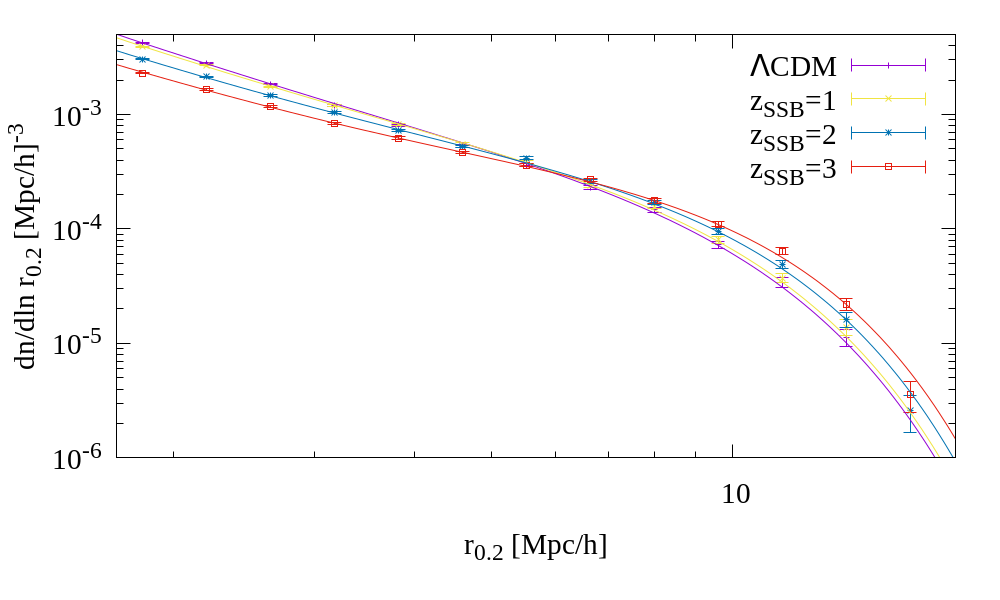}
    \caption{Similar to Fig.~\ref{fig:bias}, but for the void abundance measured and fit in $f(R)$ and symmetron simulations, showing more large (fewer small) voids in the MG scenarios.
    }
    \label{fig:recovered_abundance}
\end{figure}

\end{appendix}

\end{document}